\documentclass[10pt]{article}
\usepackage{arxiv}
\usepackage[T1]{fontenc}
\usepackage[utf8]{inputenc}
\usepackage{amsmath,amssymb,graphicx,booktabs,array,url,textcomp}
\usepackage[hidelinks]{hyperref}
\usepackage{microtype}
\usepackage{float}
\newcommand{\sourcehighlight}[1]{#1}
\newif\ifincludeAGUExtras
\includeAGUExtrasfalse

\renewcommand{\shorttitle}{}
\title{An Adaptive-Horizon Graph Neural Operator for coupled surrogate morphodynamic modelling}
\author{
\textbf{Ying Cao}$^{1}$\quad
\textbf{Yifan Yang}$^{1,*}$\quad
\textbf{Junqiang Xia}$^{1,*}$\\[4pt]
\textbf{Julia C. Mullarney}$^{2}$\quad
\textbf{M\'ario J. Franca}$^{3}$\\[9pt]
\normalsize $^{1}$ State Key Laboratory of Water Resources Engineering and Management,\\
\normalsize Wuhan University, Wuhan, China\\[3pt]
\normalsize $^{2}$ Coastal Marine Group, School of Science, The University of Waikato, Hamilton, New Zealand\\[3pt]
\normalsize $^{3}$ Institute for Water and Environment (IWU), Karlsruhe Institute of Technology (KIT), Karlsruhe, Germany
}
\date{}
\hypersetup{
  pdftitle={An Adaptive-Horizon Graph Neural Operator for coupled surrogate morphodynamic modelling},
  pdfauthor={Ying Cao, Yifan Yang, Junqiang Xia, Julia C. Mullarney, Mario J. Franca},
  pdfkeywords={graph neural operator, morphodynamics, TELEMAC-2D, GAIA, adaptive temporal bundling}
}
\begin{document}
\maketitle
\begingroup
\renewcommand{\thefootnote}{\fnsymbol{footnote}}
\footnotetext[1]{Corresponding authors: Yifan Yang (\href{mailto:yangwbpo@gmail.com}{yangwbpo@gmail.com}); Junqiang Xia (\href{mailto:xiajq@whu.edu.cn}{xiajq@whu.edu.cn}).}
\endgroup
\begin{abstract}

Morphodynamic simulations require repeated coupling between hydrodynamic solvers, sediment transport modules, and bed updating. For long duration, multi-scenario simulations on large unstructured meshes, such coupling is computationally demanding. This study proposes an Adaptive-Horizon Graph Neural Operator (AH-GNO) that formulates bed evolution as a graph operator learning problem from recent hydrodynamic and topographic states to predict future bed elevation increments. The model operates directly on the native unstructured meshes. The approach integrates area-weighted graph learning, historical state selection, and forecast horizon estimation, allowing the effective history length (\emph{K}) and update interval (\emph{B}) to vary with the evolving flow and bed state. Using Telemac2D/GAIA solutions as supervised numerical references, the framework is evaluated on the official Bump and Yen bend benchmarks and on a field-scale application (Waiho River, New Zealand). Results show that the AH-GNO reproduces the main patterns of bedform migration, bend-scale erosion and deposition, and localized bed adjustment. Across the three cases, the tested CPU-GPU deployment reduced post-training bed-update workflow time from 12\textasciitilde1204 s with GAIA to 0.08\textasciitilde3.34 s with AH-GNO. This improvement corresponds to deployment-specific accelerations of approximately 150\textasciitilde360 times relative to 4-core GAIA CPU configurations, and with 105\textasciitilde186 times gains for 20-core CPU configurations. These results suggest that, while retaining hydrodynamic feedback from TELEMAC-2D, a learned operator for bed updating can provide an efficient acceleration component for morphodynamic scenario analysis and rapid coupled simulation.
\end{abstract}

\ifincludeAGUExtras
\subsection*{Key Points}
\begin{itemize}
\item An adaptive-horizon graph neural operator learns state-dependent history length and prediction horizon for morphodynamic evolution.
\item The learned operator predicts bed-elevation changes and can be coupled online with hydrodynamic solvers or surrogate flow models.
\item Benchmark and real river tests demonstrate accurate bed-evolution prediction and order-of-magnitude wall-clock speedups.
\end{itemize}

\subsection*{Plain Language Summary}
River beds change as water erodes, transports, and deposits sediment. Numerical models can simulate these processes, but they are often too expensive for long simulations or for testing many flood and engineering scenarios. We develop a machine-learning model that learns how the river bed changes from recent information on water depth, velocity, and bed elevation, while operating directly on the original unstructured model mesh. The model adaptively decides how much past information to use and how many future bed updates should be applied before the hydrodynamic model needs to be updated again. Tests on two standard TELEMAC/GAIA benchmarks and one field-scale braided-river case show that the method captures the main erosion and deposition patterns while greatly reducing computational time. The approach is intended to accelerate morphodynamic models based on physical processes, rather than replace them.

\fi

\section{Introduction}

River-bed evolution is a fundamental process in fluvial morphodynamics with continually changing bed elevation and channel geometry under the combined effects of erosion, sediment transport, and deposition. Changes in the riverbed both respond to and influence the flow \sourcehighlight{(Paul \& Biswas, 2019)}, by modifying water depth, velocity, bed shear stress \sourcehighlight{(Deal et al., 2023)}, and flood conveyance. Reliable simulation of river-bed morphodynamics is important for river engineering, flood-risk assessment, and sediment management \sourcehighlight{(Afan et al., 2016)}.

Several mature process-based models are now used for riverine, estuarine, and coastal morphodynamic simulation. Delft3D has been widely applied to coastal and fluvial morphodynamics \sourcehighlight{(Vundavilli et al., 2021, 2024)}, and Delft3D-FM further supports three-dimensional morphodynamic modeling on flexible unstructured meshes \sourcehighlight{(Mayerle et al., 2015)}. BASEMENT provides modules for river hydrodynamics and morphodynamics, including two-dimensional formulations based on the Saint-Venant equations and Exner-type bed evolution \sourcehighlight{(Vanzo et al., 2021)}. HEC-RAS includes sediment-transport and movable-bed capabilities for engineering applications \sourcehighlight{(Foda et al., 2025)}. In the TELEMAC-MASCARET system, the GAIA module computes sediment transport and bed evolution and can be coupled with TELEMAC-2D on unstructured meshes \sourcehighlight{(Tassi et al., 2023)}. Although these models are physically interpretable and widely used, computational cost remains a major limitation in process-based morphodynamic modeling. Resolving events such as large floods, which can reorganize channel morphology over short periods \sourcehighlight{(Yu et al., 2025)}, requires repeated hydrodynamic solution, sediment-transport calculation, bed updating, wetting--drying treatment, and stability control, and this cost becomes more pronounced for large domains, refined or unstructured meshes, and realistic river geometries \sourcehighlight{(Hu et al., 2022; Reyns et al., 2026)}. Practical applications often require many repeated simulations for calibration, sensitivity analysis, uncertainty quantification, flood-scenario comparison, and engineering-design assessment, making exhaustive process-based modeling difficult for routine decision support \sourcehighlight{(Cheng et al., 2023; Kroon et al., 2025)}. Morphological acceleration factors can reduce part of the computational burden by accelerating bed evolution \sourcehighlight{(Reyns et al., 2026)}, but these factors introduce additional assumptions about the relationship between hydrodynamic forcing and morphological response, and their reliability can depend on flow unsteadiness, spatial scale, and sediment-transport regime \sourcehighlight{(Morgan et al., 2020)}.

Hydrodynamic surrogates to predict flood inundation, water depth, velocity, discharge, and reach-scale water levels are being increasingly developed. These approaches use deep neural networks, convolutional and recurrent networks, U-Nets, graph neural networks, and physics-informed neural networks to approximate processes governed by the shallow-water equations or high-fidelity two-dimensional flood models \sourcehighlight{(Bentivoglio et al., 2023; Sun et al., 2021; Taghizadeh et al., 2025)}. These efforts range from purely data-driven networks trained on the output of SWE-based hydrodynamic models \sourcehighlight{(Haces-Garcia et al., 2025)} to physics-informed formulations that embed the two-dimensional shallow-water equations directly in the learning framework \sourcehighlight{(Liu et al., 2026; Qi et al., 2024)}, to deep-learning surrogates used for bathymetry inversion of shallow-water solvers \sourcehighlight{(Liu et al., 2024)}, and to hybrid strategies that correct low-fidelity flood-inundation models through spatial analysis and Gaussian-process learning at reduced cost \sourcehighlight{(Fraehr et al., 2022)}. Recent reviews have also examined how such tools are selected, developed, and compared across flood-related geospatial analysis more broadly \sourcehighlight{(Pourzangbar et al., 2024)}.

Bed-elevation change, however, follows sediment mass conservation. In Exner-type formulations, the change equals the divergence of the sediment flux, a quantity defined by spatial gradients rather than by the hydraulic state at a single point, and the flux depends nonlinearly on water depth, velocity, bed shear stress, grain size, bed slope, and sediment supply \sourcehighlight{(Paola \& Voller, 2005)}. This constraint makes bed evolution harder to predict than single-step water levels or point-scale sediment transport. For sediment transport and morphodynamics, machine learning has more often targeted bed-load transport rate, suspended-sediment load, sediment concentration, sediment yield, and local scour depth, using approaches that span machine-learning regression models for open-channel sediment transport \sourcehighlight{(Abid Almubaidin et al., 2023)}, large-sample data-driven and ensemble or tree-based regressors for river bed load \sourcehighlight{(Hosseiny et al., 2023; Mir et al., 2024)}, and physics-informed networks that encode sediment-transport dynamics for pipe-flow prediction \sourcehighlight{(Tipu et al., 2025)}. Across these studies, sensitivity and SHAP (SHapley Additive exPlanations) analyses have repeatedly identified dimensionless bed shear stress and the Shields parameter as dominant predictive variables. Reviews of the wider field report the same emphasis on transport rates and point-scale morphological indicators rather than on distributed bed-elevation fields \sourcehighlight{(Goldstein et al., 2019)}.

The data-driven and physics-informed models capture nonlinear relations among hydraulic and sediment variables and are effective for local parameterization, empirical-formula replacement, or offline component approximation. Their targets are scalar quantities at a point or a cross section, and they generally do not operate as mesh-based bed-update modules inside the time-stepping loop of a fully coupled hydro-morphodynamic solver \sourcehighlight{(Fathi et al., 2026)}. If the bed is updated independently of the hydrodynamic solver, the predicted trajectory can drift from the physically constrained path because the surrogate is repeatedly evaluated on states affected by the cumulative errors \sourcehighlight{(Kochkov et al., 2021)}. A more appropriate strategy is therefore not to replace the process model outright, but to embed the surrogate within a coupled workflow as an accelerated bed update, while the solver continues to provide hydrodynamic feedback and preserve the coupled flow--sediment--bed evolution structure.

Neural-operator learning provides an effective framework for this kind of spatially distributed surrogate, rather than conventional regression. A standard neural network, (e.g., a multilayer perceptron, a convolutional network, or many graph regressors), learns a mapping between finite-dimensional vector spaces that is tied to a fixed discretization, so inputs and outputs are given on a specific grid or node set. Thus, changing the mesh resolution generally requires retraining. A neural operator instead approximates a mapping between infinite-dimensional function spaces, that is, the solution operator of a partial differential equation, so that the learned object or variable is largely independent of the discretization. So in principle, results can be generalized across mesh resolutions \sourcehighlight{(Kovachki et al., 2023)}. DeepONet introduced a neural-network framework for learning nonlinear operators between function spaces \sourcehighlight{(Lu et al., 2021)}, whereas the Fourier Neural Operator learns solution operators for parameterized partial differential equations and has shown high efficiency for problems such as the Burgers equation, Darcy flow, and Navier-Stokes flow \sourcehighlight{(Li et al., 2021)}. The graph kernel neural operator approximates the underlying integral operator through area-weighted neighborhood aggregation on irregular point sets \sourcehighlight{(Li et al., 2020)}. Operator-learning approaches have since been extended to latent-space formulations for real-time prediction of complex dynamics \sourcehighlight{(Kontolati et al., 2024)} and to physics-informed operators for high-fidelity flow simulation \sourcehighlight{(Zhao et al., 2025)}. For river morphodynamics, the target is not a single sediment variable but the spatially distributed bed-elevation update over irregular domains and unstructured meshes. Graph-based models are therefore particularly suitable \sourcehighlight{(Li et al., 2023; Mousavi et al., 2025)}. MeshGraphNets showed that graph neural networks can learn mesh-based physical simulations through mesh connectivity \sourcehighlight{(Pfaff et al., 2021)}. GraphDeepONet further combines graph neural networks with operator-learning concepts for time-dependent partial differential equations on irregular meshes and arbitrary spatial points \sourcehighlight{(Cho et al., 2026)}. These methods can naturally represent numerical meshes, boundary information, upstream-downstream connectivity, and spatially varying hydrodynamic state.

Existing time-dependent neural operators usually employ fixed temporal settings: a fixed number of historical states as input and a fixed number of future states predicted per inference step, as in temporal-bundling strategies that forecast several steps together to reduce solver calls and error accumulation \sourcehighlight{(Brandstetter et al., 2022; Phillip Lippe et al., 2023)}. This assumption is not well suited to river morphodynamics, where the effective history length and reliable forecast horizon vary with the current flow and bed state. During rapid scour, bedform migration, channel adjustment, or strong flood forcing, the model needs a shorter prediction interval and more frequent hydrodynamic updates; during slow or near-equilibrium evolution, a longer horizon may be both reliable and computationally advantageous. A fixed window thus imposes the same time scale on different evolutionary states, wasting computation during stable periods and risking unreliable extrapolation during rapid adjustment, which motivates making the forecast horizon state-dependent rather than a fixed numerical setting.

This study develops an Adaptive-Horizon Graph Neural Operator (AH-GNO) for simulating river-bed evolution on unstructured meshes and integrates the operator into a conventional 2D hydro-morphodynamic modeling workflow (TELEMAC-2D and GAIA). The basic idea is illustrated in Figure 1. Bed evolution is formulated as a graph-operator learning problem, allowing bed-elevation changes to be predicted directly on the native computational mesh. An adaptive temporal mechanism is introduced to infer the effective historical contribution and prediction horizon, replacing the fixed temporal window commonly used in time-dependent surrogate models. The learned operator is designed to be coupled with TELEMAC-2D/GAIA, with the hydrodynamic solver retained in the loop and GAIA morphodynamic solutions used for training and validation. The framework is evaluated using two official TELEMAC/GAIA benchmark cases, Bump and Yen, and a field-scale braided-river application on the Waiho River, to assess computational efficiency, bed-prediction accuracy, and long-rollout stability.AH-GNO is assessed against a non-adaptive one-step counterpart built on the same graph operator, so that the contribution of the adaptive temporal mechanism can be isolated.

\begin{figure}[tbp]
\centering
\includegraphics[width=\linewidth,height=0.72\textheight,keepaspectratio]{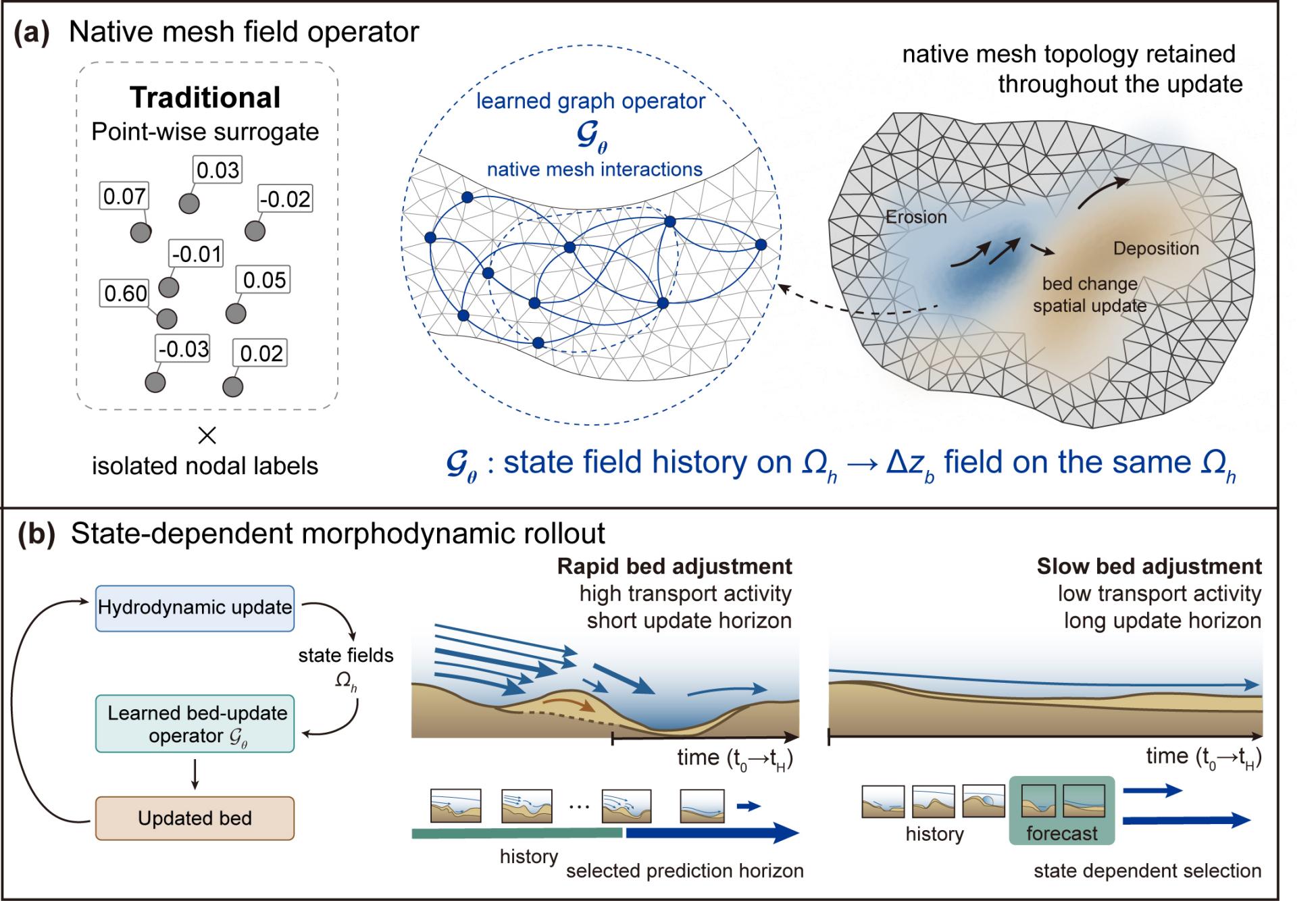}
\caption{Conceptual motivation for the native-mesh graph operator and adaptive-horizon morphodynamic surrogate. (a) Bed evolution is treated as a field-operator problem on the native unstructured mesh, rather than as an independent point-wise prediction. (b) River-bed adjustment evolves at different temporal rates; the surrogate selects the effective history and prediction horizon according to the bed-change process, using shorter horizons for rapid adjustment and longer horizons for slow adjustment.}
\label{fig:1}
\end{figure}

\section{Adaptive-Horizon Graph Neural Operator}

\subsection{Reference morphodynamic solver: TELEMAC-2D/GAIA}

The reference system used in this study is the open-source TELEMAC-MASCARET suite, in which TELEMAC-2D computes the depth-averaged time-dependent free-surface flow, and the GAIA module computes sediment transport and the resulting bed evolution. Water depth and the two depth-averaged velocity components are defined at the nodes of an unstructured triangular mesh, which TELEMAC-2D uses to integrate the shallow-water equations. The hydrodynamic continuity equation can be written as:

\begin{equation}
\frac{\partial h}{\partial t} + \nabla \bullet (h\mathbf{u}) = 0,
\label{eq:1}
\end{equation}

where \(h\) is the water depth and \(\mathbf{u} = (u,v)\) are the depth-averaged velocity components in \emph{x} and \emph{y} directions, respectively. The momentum equations are solved on the same mesh and provide the velocity field and bed shear stress that drive sediment transport.

Bed evolution in GAIA is governed by an Exner equation:

\begin{equation}
(1 - p)\frac{\partial z_{b}}{\partial t} + \nabla \cdot \mathbf{q}_{s} = 0,
\label{eq:2}
\end{equation}

where \(z_{b}\) is the bed elevation, \(p\) is the bed porosity, and \(\mathbf{q}_{s}\) is the volumetric sediment-transport flux. At each morphodynamic time step, GAIA evaluates \(\mathbf{q}_{s}\) from the local hydraulic state using a sediment transport closure, and the Exner equation is then integrated to update \(z_{b}\). This sequence, including hydrodynamic solution, transport-flux evaluation, and bed update, is repeated over the simulation, and these steps constitute the principal computational cost of long morphodynamic runs.

The objective of this work is not to replace the entire TELEMAC-2D/GAIA system but to replace only the bed-evolution update. Rather than computing \(\mathbf{q}_{s}\) explicitly and integrating the Exner equation, we train an operator to map a short history of the coupled hydrodynamic--bed state to the subsequent bed-elevation increments. Let \(\mathbf{s} = (x,y)\) denote a spatial coordinate. At a sampled time t, the hydro-morphodynamic state is

\begin{equation}
\mathbf{X}_{t} = \left( u_{t},\, v_{t},\, h_{t},\, z_{b,t} \right).
\label{eq:3}
\end{equation}

Given a finite history of such states, the objective is to learn a parametric operator \(\mathcal{G}_{\theta}\) that maps recent flow-bed states to future bed-elevation increments,

\begin{equation}
\mathcal{G}_{\theta}:\ \left( \mathbf{X}_{t - K + 1},\,\ldots,\,\mathbf{X}_{t} \right)\  \rightarrow \ \left( \Delta z_{t + 1},\,\ldots,\,\Delta z_{t + B} \right),
\label{eq:4}
\end{equation}

where the effective history length K and the write-back horizon B are not fixed but are inferred from the current state (further discussion on the adaptive history and horizon window is given in Sections 2.3--2.4). Write-back denotes the operation of returning predicted bed-elevation increments to the TELEMAC-2D bed field, and B is the number of consecutive predicted increments applied before the hydrodynamic solver is called again. The target is the bed-elevation increment \(\Delta z_{b}\) rather than the absolute elevation. This choice isolates the morphodynamic update over a coupling interval and reduces the tendency of the model to reproduce static bathymetric structure. During online rollout, the predicted increments are accumulated and written back to the TELEMAC bed elevation. The learned map is therefore a surrogate for the sediment transport/Exner update, not for the full hydro-morphodynamic system.

\subsection{Graph formulation and input-output representation}

The model operates directly on the native unstructured triangular mesh of TELEMAC-2D and requires no interpolation to a regular grid. The computational mesh is represented as a graph:

\begin{equation}
\mathcal{M} =\left( \mathcal{V,E} \right),
\label{eq:5}
\end{equation}

where \(\mathcal{V}\) is the set of mesh nodes and \(\mathcal{E}\) denotes the neighborhood relations used for aggregation. The mesh contains \(N\) nodes, and the coordinate of node \(i\) is \(\mathbf{s}_{i} = \left( x_{i},y_{i} \right)\). The continuous state \(\mathbf{X}_{t}\) is discretized at the mesh nodes as

\begin{equation}
\mathbf{X}_{i,t} = \left( u_{i,t},\, v_{i,t},\, h_{i,t},\, z_{b,i,t} \right),i = 1,\ldots,N.
\label{eq:6}
\end{equation}

The velocity components describe flow intensity and direction, water depth controls transport capacity and wetting-drying state, and bed elevation defines the evolving topographic boundary that further influences subsequent flow and sediment response.

To make graph aggregation and error evaluation comparable across different mesh densities, a lumped control area is assigned to each node. For a triangular element \emph{T}, the control area associated with node \emph{i} is defined as

\begin{equation}
A_{i} = \sum_{T \ni i}^{}\frac{|T|}{3},
\label{eq:7}
\end{equation}

where \emph{\textbar T\textbar{}} is the physical area of triangle \emph{T} and \(T \ni i\) denotes all triangular elements that contain node \emph{i}. This area weight is used both as a numerical quadrature weight in the graph integral operator and in the area-weighted training loss and evaluation metrics. Compared with simple node averaging, the area-weighted form more closely approximates integration over the physical domain and reduces the influence of nonuniform mesh density or local mesh refinement on training and performance assessment.

Let \(K_{\max}\) denote the maximum history length and \(\mathrm{\Delta}t_{b}\) the base sampling interval. At time \emph{t}, the discrete input window consists of \(K_{\max}\) states is

\begin{equation}
\mathcal{X}_{t} = \left\{ \mathbf{X}_{t - \left( K_{\max} - 1 \right)\Delta t_{b}},\,\ldots,\,\mathbf{X}_{t - \Delta t_{b}},\,\mathbf{X}_{t} \right\}.
\label{eq:8}
\end{equation}

At each node, the four physical variables from the \(K_{\max}\) historical states are normalized and concatenated as model input features. The supervised learning target is a sequence of future bed-elevation increments,

\begin{equation}
\mathbf{Y}_{i,t} = \left( \Delta z_{b,i,t + 1},\,\ldots,\,\Delta z_{b,i,t + B_{\max}} \right),
\label{eq:9}
\end{equation}

where the \emph{j}-th future increment is

\begin{equation}
\mathrm{\Delta}z_{b,i,t + j} = z_{b,i,t + j\mathrm{\Delta}tb} - z_{b,i,t + (j - 1)\mathrm{\Delta}tb}.
\label{eq:10}
\end{equation}

All normalization statistics for inputs and targets are computed only from the training set and are estimated using area-weighted means and standard deviations. This technique avoids leakage of validation or test information through the normalization parameters and makes the statistics representative of the physical domain rather than of the number of mesh nodes.

\subsection{Adaptive-Horizon Graph Neural Operator (AH-GNO)}

The architecture follows an encode--process--decode design (see Figure 2). A pointwise lifting network encodes each historical nodal state, a temporal selector fuses the history into a single latent field, a stack of graph integral operators models spatial interactions, and two decoding heads map the resulting latent field to the bundled bed-elevation increments and to the write-back horizon. The graph integral operator is a discrete realization of the graph kernel or neural-operator formulation of Li et al. (2020) and is implemented with the neuraloperator library (Kovachki et al., 2023); the temporal selector and horizon head are introduced here to make the history length and forecast horizon state-dependent.

\textbf{Lifting.} Each historical nodal state is mapped into a high-dimensional latent space by a shared pointwise network. Denoting normalized variables with a hat, the lifted representation of node \(i\) at historical index \(k\) is

\begin{equation}
\mathbf{e}_{i,k} = \phi_{lift}\left( {\widehat{u}}_{i,k},\,{\widehat{v}}_{i,k},\,{\widehat{h}}_{i,k},\,{\widehat{z}}_{b,i,k} \right) \in \mathbb{R}^{d},
\label{eq:11}
\end{equation}

where \(\phi_{lift}\) is a shared multilayer perceptron and \(d\) is the latent width. The same network is applied to all nodes and all historical states, so the model parameters do not depend on a fixed node index or mesh size.

\textbf{Adaptive historical selection.} Conventional time-dependent neural operators consume a fixed number of historical states. Here, \(K_{\max}\) is treated only as an upper bound, and a learned selector decides how much history to use for each prediction. A global descriptor of the most recent state is first formed by an area-weighted average,

\begin{equation}
\mathbf{g}_{K} = \frac{\,\sum_{i}^{}A_{i}\mathbf{e}_{i,K_{\max}}}{\sum_{i}^{}A_{i}},
\label{eq:12}
\end{equation}

which \(\mathbf{g}_{K}\) represents a global summary of the current hydrodynamic--topographic field while remaining insensitive to node density. A small network maps this descriptor to raw temporal weights,

\begin{equation}
{\widetilde{\alpha}}_{k} = softmax(f_{K}\left( \mathbf{g}_{K} \right))_{k},\ k = 1,\ldots,K_{\max}.
\label{eq:13}
\end{equation}

where \(\alpha_{k}\) denotes the contribution of the k-th historical state to the current prediction. To sharpen temporal selection, weights below a fraction \(\rho\) of the uniform value are discarded and the remainder is renormalized,

\begin{equation}
\alpha_{k} = \frac{{\widetilde{\alpha}}_{k}\mathbb{1}\left\lbrack {\widetilde{\alpha}}_{k} \geq \rho/K_{\max} \right\rbrack}{\sum_{m}^{}{\widetilde{\alpha}}_{m}\mathbb{1}\left\lbrack {\widetilde{\alpha}}_{m} \geq \rho/K_{\max} \right\rbrack},
\label{eq:14}
\end{equation}

where \(\mathbb{1}\lbrack \cdot \rbrack\) is the indicator function (one when its condition holds and zero otherwise) and \(\rho\) is a relative threshold factor, set to \(\rho = 0.5\); if the threshold removes all weights, the single largest weight is retained. The sparsification threshold is therefore relative to the uniform distribution \(1/K_{\max}\) rather than an absolute value. The hard masking is non-differentiable, so a straight-through estimator is used: the forward pass uses the sparse weights \(\alpha_{k}\), while gradients flow through the dense weights \({\widetilde{\alpha}}_{k}\),

\begin{equation}
\alpha_{k}^{ste} = {\widetilde{\alpha}}_{k} + sg\left( \alpha_{k} - {\widetilde{\alpha}}_{k} \right),
\label{eq:15}
\end{equation}

where \(sg( \cdot )\) denotes the stop-gradient operator. The selected weights fuse the history into the initial latent field of the spatial operator,

\begin{equation}
\mathbf{h}_{i}^{(0)} = \sum_{k = 1}^{K_{\max}}\alpha_{k}^{ste}\,\mathbf{e}_{i,k}.
\label{eq:16}
\end{equation}

Physically, rapid scour or deposition tends to concentrate weight on the most recent states, whereas slowly evolving periods can retain a longer memory; the effective history length \(K_{eff} = \sum_{k}^{}k\alpha_{k}\) thus varies with the flow--bed state.

\textbf{Spatial graph integral operator.} Spatial interactions are represented by \(L\) layers of an area-weighted graph integral operator. For node \(i\), the neighborhood \(N(i) = \{ j: \parallel \mathbf{s}_{i} - \mathbf{s}_{j} \parallel \leq r\}\) is defined by a fixed search radius \(r\) in normalized coordinates---the node coordinates are rescaled to the unit bounding box \({\lbrack 0,1\rbrack}^{2}\) of the domain, so that r represents a fixed fraction of the domain extent---and the number of neighbors adapts to local mesh density. The layer update is

\begin{equation}
\mathbf{h}_{i}^{(l + 1)} = \mathbf{h}_{i}^{(l)} + \sigma\left( \frac{\sum_{j \in N(i)}^{}A_{j}\kappa_{\theta}^{(l)}\left( \mathbf{s}_{i} - \mathbf{s}_{j} \right) \odot \mathbf{h}_{j}^{(l)}}{\sum_{j \in N(i)}^{}A_{j}} \right),
\label{eq:17}
\end{equation}

where \(\sigma\) is a nonlinear activation, \(\odot\) is the elementwise product, and \(\kappa_{\theta}^{(l)}\) is a learnable channelwise kernel. The kernel depends only on the relative coordinate, encoded by a sinusoidal positional-embedding map \(\varphi_{pe}( \cdot )\) that lifts the two-dimensional offset into a higher-dimensional Fourier feature,

\begin{equation}
\kappa_{\theta}^{(l)}\left( \mathbf{s}_{i} - \mathbf{s}_{j} \right) = {MLP}^{(l)}\left( \left( \mathbf{s}_{i} - \mathbf{s}_{j} \right) \right),
\label{eq:18}
\end{equation}

so that the operator encodes only local geometry and is independent of absolute position or node ordering. Two design choices make the operator a discrete approximation of a continuous integral rather than ordinary message passing. First, each neighbor message is weighted by its control area \(A_{j}\), which acts as a quadrature measure. Second, the aggregated message is divided by the local area \(\sum_{j \in N(i)}^{}A_{j}\), so that the output is a local area-weighted average whose magnitude depends on the field within the search radius \(r\) rather than on the number of neighbors. The corresponding continuous form is

\begin{equation}
\frac{1}{\left| \Omega_{i} \right|}\int_{\Omega_{i}}^{}\kappa_{\theta}\left( \mathbf{s}_{i} - \mathbf{s} \right)\mathbf{h}\left( \mathbf{s} \right)d\mathbf{s}\  \approx \ \frac{\sum_{j \in N(i)}^{}A_{j}\kappa_{\theta}\left( \mathbf{s}_{i} - \mathbf{s}_{j} \right)\mathbf{h}_{j}}{\sum_{j \in N(i)}^{}A_{j}},
\label{eq:19}
\end{equation}

where \(\Omega_{i}\) is the local interaction region around node \(i\) and \(\left| \Omega_{i} \right| = \sum_{j \in N(i)}^{}A_{j}\). This local area normalization decouples the operator from node density and is the mechanism that gives the model its discretization-consistent behavior across meshes. This feature thus distinguishes the present graph neural operator approach from the use of mesh-topology-bound graph neural networks (see Section 4).

\textbf{Bundled increment decoding.} After the final graph integral layer, a pointwise projection multilayer perceptron \(Q_{\theta}\) maps the latent state at each node to \(B_{\max}\) future bed-elevation increments,

\begin{equation}
\mathbf{Y}_{i,t} = Q_{\theta}\left( \mathbf{h}_{i}^{(L)} \right) \in \mathbb{R}^{B_{\max}}.
\label{eq:20}
\end{equation}

\textbf{Forecast-horizon head.} In addition to the full increment sequence, a horizon head estimates how many of the predicted increments can be reliably written back to TELEMAC-2D under the current state. A global descriptor is formed from the final latent field,

\begin{equation}
\mathbf{g}_{B} = \frac{\sum_{i}^{}A_{i}\mathbf{h}_{i}^{(L)}}{\sum_{i}^{}A_{i}},
\label{eq:21}
\end{equation}

and a classifier network \(f_{B}\) (a multilayer perceptron) produces a distribution over candidate horizons \(H \in \{ 1,\ldots,B_{\max}\}\),

\begin{equation}
\mathbf{p}_{B} = softmax\left( f_{B}\left( \mathbf{g}_{B} \right) \right),p_{B,h} = P(H = h).
\label{eq:22}
\end{equation}

To obtain a per-step reliability that is monotone by construction, the horizon distribution is converted into cumulative survival weights,

\begin{equation}
\beta_{j} = \sum_{b = j}^{B_{\max}}p_{B,b} = P(H \geq j),j = 1,\ldots,B_{\max},
\label{eq:23}
\end{equation}

so that \(\beta_{1} \geq \beta_{2} \geq \ldots \geq \beta_{B_{\max}}\). The effective write-back length is read off from these weights, with at least one increment applied at every coupling call,

\begin{equation}
B_{eff} = max\left( 1,\ |\{ j:\beta_{j} > \frac{1}{2}\}| \right).
\label{eq:24}
\end{equation}

Decoupling the predicted increment sequence from the decision of how many increments are written back allows the model to reduce bed-update calls during slowly evolving periods while avoiding unstable long extrapolation during rapid adjustment. The supervision used to train this head is defined in Section 2.5.

\begin{figure}[tbp]
\centering
\includegraphics[width=\linewidth,height=0.72\textheight,keepaspectratio]{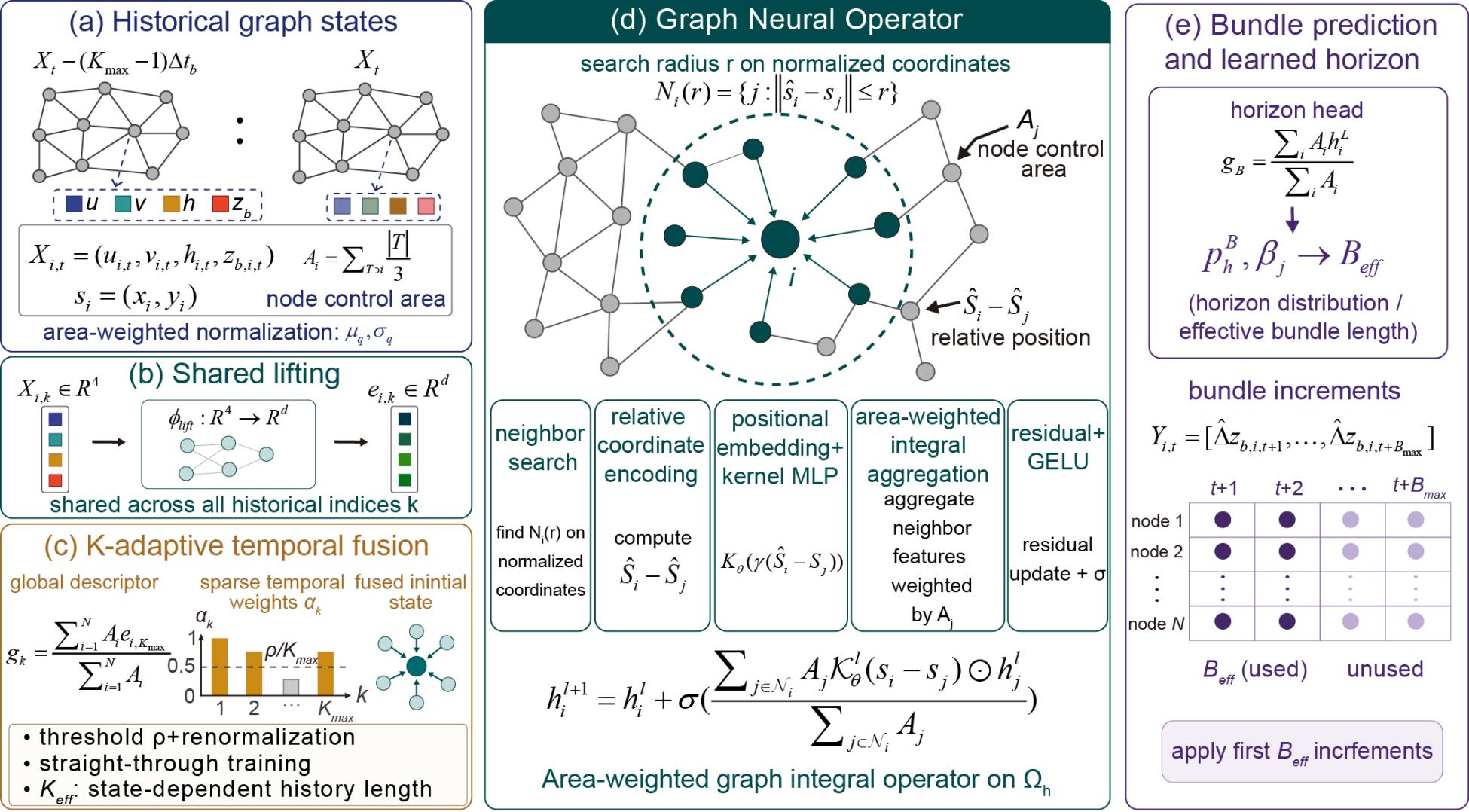}
\caption{Architecture of the Adaptive-Horizon Graph Neural Operator (AH-GNO).}
\label{fig:2}
\end{figure}

\subsection{Adaptive temporal bundling method}

Brandstetter introduced temporal bundling for graph-based PDE solvers, in which several future states are predicted synchronously at each inference call rather than advanced one step at a time \sourcehighlight{(Brandstetter et al., 2022)}. This strategy reduces the number of solver calls and helps mitigate error accumulation during autoregressive rollout. Cho subsequently adopted this idea with a fixed bundle length \sourcehighlight{(Cho et al., 2026)}. In these formulations both the history length and the bundle size are prescribed constants. This assumption is not well suited to river morphodynamics, where the informative history and the reliable forecast range depend on the current state: during rapid scour, bedform migration, or strong flood forcing, a shorter horizon and more frequent hydrodynamic updates are required, whereas during slow or near-equilibrium evolution a longer horizon is both reliable and efficient. AH-GNO therefore generalizes fixed temporal bundling into a state-dependent form, in which the effective history length and the write-back horizon are inferred online rather than fixed.

The two adaptive mechanisms are realized by the components defined in Section 2.3 and are not repeated here. The historical selector (the sparse-softmax weights \(\alpha_{k}\)) determines the effective history \(K_{eff}\) fused into the graph operator, so that \(K_{\max}\) acts only as an upper bound on memory. The forecast-horizon head (the survival weights \(\beta_{j}\) and the effective length \(B_{eff}\)) determines how many bundled increments are written back during the current coupling call. Together they form a state-dependent temporal-bundling policy: longer bundled updates are used when bed evolution is smooth and predictable, and the write-back horizon is shortened when rapid adjustment, localized scour, or wetting--drying transitions raise the extrapolation risk.

\subsection{Loss function}

Model training uses a composite loss defined on normalized bed-elevation increments, combining an area-weighted prediction term with a horizon-supervision term. Let \({\widehat{y}}_{ij}\) and \(y_{ij}\) denote the predicted and reference normalized increments at node \(i\) and future step \(j\). To focus learning on active erosion and deposition regions, an activity weight is assigned to each node:

\begin{equation}
\omega_{i} = 1 + \gamma clip\left( \frac{\max_{j}\left| \Delta z_{b,ij} \right|}{s_{a}},0,1 \right),
\label{eq:25}
\end{equation}

where \(clip( \cdot ,0,1)\) clamps its argument to the interval \(\lbrack 0,1\rbrack\), \(s_{a}\) is a bed-activity scale estimated from the training data, and \(\gamma > 0\) is a fixed gain. Nodes with appreciable bed change within the prediction window receive larger weight; nearly stable nodes keep \(\omega_{i} \approx 1\).

\textbf{Prediction loss.} The main term is an area-weighted Huber loss expressed as:

\begin{equation}
\mathcal{L}_{pred} = \frac{\sum_{i}^{}{\sum_{j}^{}A_{i}}\omega_{i}\rho_{\delta}\left( {\widehat{y}}_{ij} - y_{ij} \right)}{B_{\max}\sum_{i}^{}A_{i}\omega_{i}},
\label{eq:26}
\end{equation}

where \(\rho_{\delta}\) is the Huber penalty:

\begin{equation}
\rho_{\delta}(e) = \left\{ \begin{matrix}
\frac{e^{2}}{2\delta}, & |e| < \delta, \\
|e| - \frac{\delta}{2}, & |e| \geq \delta.
\end{matrix} \right.\
\label{eq:27}
\end{equation}

The Huber penalty is used because bed-elevation-increment learning contains both many small residuals over weakly active areas and sparse, localized large residuals near scour fronts, deposition margins, wetting-drying interfaces, and boundaries. A purely squared-error loss would give these few extreme entries disproportionately large gradients and can make training sensitive to local numerical or phase errors, whereas a purely absolute-error loss is less smooth near zero and gives weaker incentives to refine small physically meaningful errors. The Huber penalty provides a compromise: it is quadratic for \(|e| \leq \delta\), which preserves the local accuracy and smooth gradients of an L2 loss for ordinary residuals, and it becomes linear for \(|e| > \delta\), so the influence of localized extreme residuals grows only proportionally rather than quadratically. The penalty is convex and continuously differentiable as a function of the residual e (although the full neural-network optimization remains non-convex in the model parameters), and any non-negative area/activity weighting preserves convexity with respect to each residual term. These properties make the Huber penalty suitable for constructing a stable bed-evolution operator that must learn the dominant distributed morphodynamic response without being dominated by a small number of rapidly changing nodes. We use $\delta$ = 0.20 in normalized increment units. Weighting each node by \(A_{i}\) makes the objective approximate the bed-change error integrated over the physical domain rather than a node-count average that would over-weight densely meshed regions.

\textbf{Horizon supervision.} The forecast-horizon head cannot be trained from the prediction loss alone, because the write-back length does not appear in \(\mathcal{L}_{pred}\). We therefore supervise the training process with an automatically generated pseudo-label \(H^{*}\) that records how far the bundled prediction stays within a tolerance of the reference. Using denormalized increments, the area-weighted step error and a per-step reference scale are

\begin{equation}
\varepsilon_{j} = \sqrt{\frac{\sum_{i}^{}A_{i}({\widehat{z}}_{ij} - z_{ij})^{2}}{\sum_{i}^{}A_{i}}},r_{j} = \sqrt{\frac{\sum_{i}^{}A_{i}z_{ij}^{2}}{\sum_{i}^{}A_{i}}},
\label{eq:28}
\end{equation}

where \({\widehat{z}}_{ij}\) and \(z_{ij}\) are the predicted and reference physical increments at step \(j\). A step is accepted when the error is below a relative-plus-absolute tolerance expressed as:

\begin{equation}
g_{j}= \mathbb{1}\left\lbrack \varepsilon_{j} \leq max(\eta r_{j},\ \epsilon_{0}) \right\rbrack,
\label{eq:29}
\end{equation}

where \(g_{j}\) is the per-step acceptance indicator, \(\eta\) is a relative tolerance, and \(\epsilon_{0}\) is an absolute floor; we use \(\eta = 0.5\) and \(\epsilon_{0} = 2 \times 10^{- 5}m\). The pseudo-label is the length of the longest reliable prefix:

\begin{equation}
H^{*} = clip\left( \sum_{j = 1}^{B_{\max}}\ \prod_{m = 1}^{j}g_{m},\ 1,\ B_{\max} \right),
\label{eq:30}
\end{equation}

so that \(H^{*}\) counts leading steps that are all accepted. The label is computed under stop-gradient, and the head is trained by a cross-entropy term against \(H^{*}\)that can be expressed as:

\begin{equation}
\mathcal{L}_{hor} = - logp_{B,H^{*}}.
\label{eq:31}
\end{equation}

\textbf{Composite objective.} The total loss is:

\begin{equation}
\mathcal{L} =\mathcal{L}_{pred} + \lambda_{h}\mathcal{L}_{hor},
\label{eq:32}
\end{equation}

with \(\lambda_{h} = 0.2\). Supervising the horizon distribution against \(H^{*}\) provides the gradient that the survival weights \(\beta_{j}\) would otherwise lack, while the stop-gradient label and the straight-through temporal weights (Section 2.3) keep the prediction and horizon objectives from interfering. Model selection is based primarily on the validation prediction loss, together with stability under mesh-extrapolation cases, so that the final model remains effective not only on the training mesh but also on denser or more irregular meshes.

\subsection{Online coupling}

In the online coupled simulation, TELEMAC-2D remains responsible for advancing the hydrodynamic state. At each hydrodynamic time step, the solver computes water depth and velocity over the current bed. When the base sampling interval \(\Delta t_{b}\) is reached, the nodal values of \(u\), \(v\), \(h\), and \(z_{b}\) are appended to a history buffer. Once the buffer holds \(K_{\max}\) sampled states, the normalized sequence is passed to AH-GNO, which outputs the bundled future increments together with the effective write-back length \(B_{eff}\).

The model output is transformed back to physical units and accumulated over the first \(B_{eff}\) steps. The bed elevation at node \(i\) is updated as

\begin{equation}
z_{b,i}^{new} = z_{b,i} + \sum_{j = 1}^{B_{eff}}\Delta{\widehat{z}}_{b,i,t + j},
\label{eq:33}
\end{equation}

and the water depth is made consistent with the new bed and the current free-surface elevation \(S_{i}\),

\begin{equation}
h_{i}^{new} = S_{i} - z_{b,i}^{new}.
\label{eq:34}
\end{equation}

TELEMAC-2D then continues advancing the hydrodynamics over the updated topography. In this way, the hydrodynamic solver is always retained in the loop, so AH-GNO is not an open-loop neural time-series generator but a bed-evolution component embedded within the TELEMAC-2D computation. The idea of the online coupling workflow is illustrated in Figure 3. The hydrodynamic state continues to respond to the predicted bed changes, maintaining feedback between flow and bed evolution.

\begin{figure}[tbp]
\centering
\includegraphics[width=\linewidth,height=0.72\textheight,keepaspectratio]{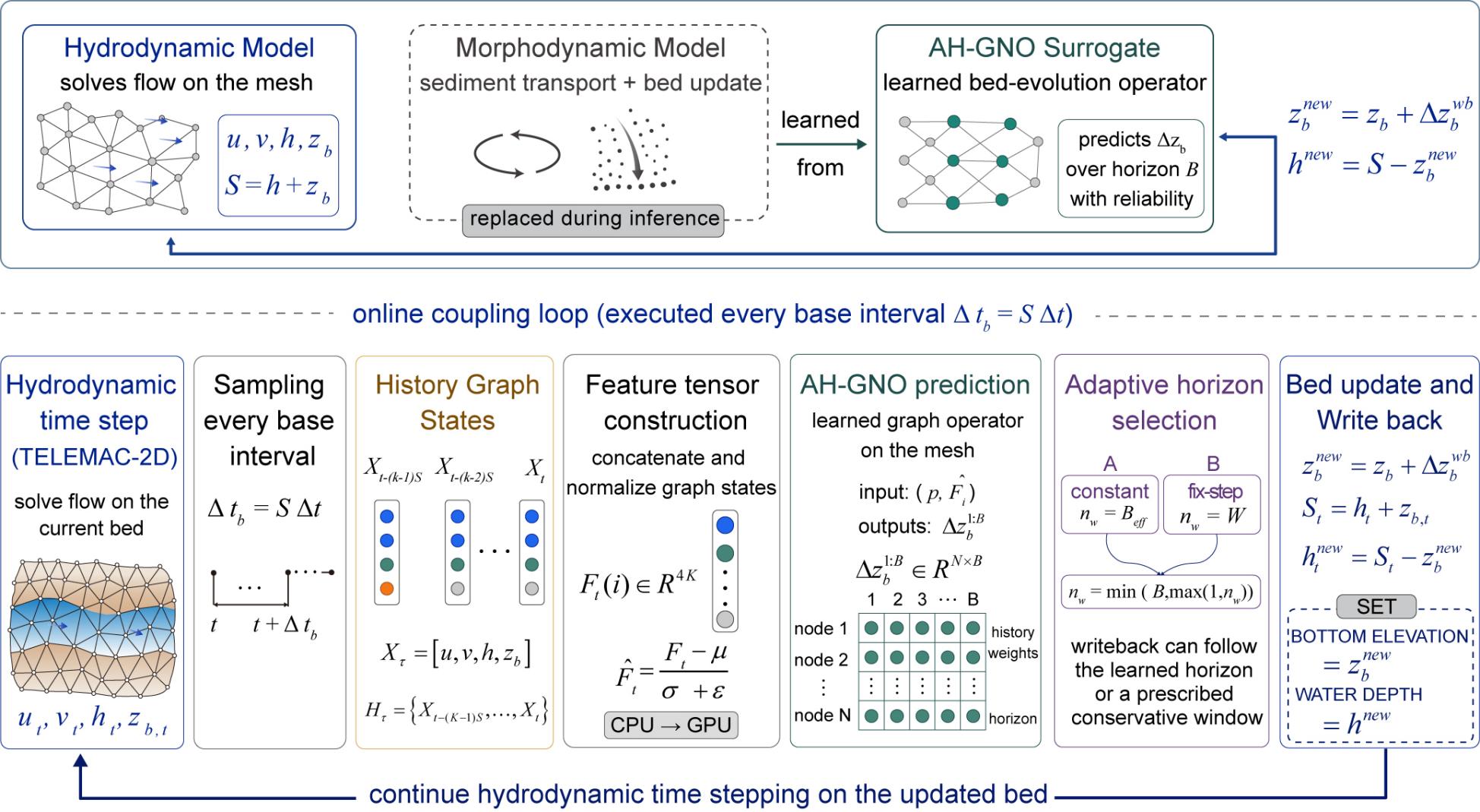}
\caption{Online coupling workflow between TELEMAC-2D and AH-GNO.}
\label{fig:3}
\end{figure}

\subsection{Implementation details}

The coupled framework is implemented in a Linux environment. TELEMAC-MASCARET version V8P4R0 is used, and the Python interface reads and writes hydrodynamic and bed-evolution fields in Selafin format. During coupling, TELEMAC-2D provides water depth, velocity, free-surface elevation, and related variables, while AH-GNO predicts the bed-elevation change exchanged back to TELEMAC-2D. AH-GNO is implemented in Python 3.12 and PyTorch 2.5.1 under Ubuntu 22.04 with the CUDA 12.4 runtime. The graph integral layers build on the neuraloperator library \sourcehighlight{(Kossaifi et al., 2026)}, using a neighbor search, channel multilayer perceptrons, and sinusoidal positional embeddings for kernel integration, feature transformation, and relative-position encoding on unstructured meshes. Preprocessing and postprocessing use NumPy; intermediate samples are stored as pickle files; results are visualized with Matplotlib; and progress is logged with tqdm. Production runs were performed on a GPU server with 20 allocated CPU cores from an Intel Xeon Platinum 8470Q processor, 90 GB of system memory, and an NVIDIA GeForce RTX 4090 GPU.

\section{Benchmark Design and Evaluation Results}

\subsection{Benchmark cases, scenario design, and data construction}

Two official TELEMAC-2D/GAIA validation cases and one field-scale braided river case (Waiho River) are selected to represent three classes of morphodynamic problems, including idealized bedform migration, unsteady bend-scale erosion and deposition, and complex real-river bed evolution (Figure 4). The Bump case is an idealized numerical test taken from the GAIA validation manual, driven by a constant inflow and a constant downstream water level, and uses an existing analytical diffusion angle as an independent reference \sourcehighlight{(Hudson \& Sweby, 2005; Vriend, 1987)}. This case is used to study riverbed form migration and lateral diffusion under simple steady-state hydraulic driving conditions. The Yen 180\textdegree{} bend case also comes from the validation manual; it reproduces the results of the bend flume experiments published by \sourcehighlight{Yen and Lee (1995)} and includes corresponding measured cross-sectional riverbed profile data; This case is driven by a transient hydrograph and is used to evaluate the model's ability to reproduce two-dimensional erosion and sedimentation patterns caused by channel curvature, curvature-induced secondary flows, and lateral sediment redistribution. The Waiho River case study is a field-scale application of a braided channel in New Zealand \sourcehighlight{(Wu et al., 2026)}, based on surveyed topographic data and a TELEMAC-2D/GAIA configuration calibrated by, and run under steady-state flow and water level conditions. All cases use validated TELEMAC-2D/GAIA numerical simulations as reference data, and AH-GNO predicts bed-elevation increments directly on the same native unstructured meshes without raster interpolation. Detailed hydrographs, water-level settings, sediment parameters, and boundary conditions are provided in the Supporting Information; the main text focuses on model training, rollout, prediction accuracy, and computational cost.

Scenario splits are performed at the case level (see Table 1). A scenario is one complete TELEMAC-2D/GAIA simulation with its own boundary forcing, and each scenario is assigned in its entirety to the training, validation, or test set, so that no output frame of a test scenario is seen during training. For the Bump case, eight steady-state scenarios are generated: 6 for training, 1 for validation, and 1 for testing. Each Bump scenario is simulated for 100 h, which under the prescribed Grass transport law is the time the conical dune needs to migrate far enough downstream for its path and its lateral spreading angle to be measured reliably, and bed and hydrodynamic fields are saved every 10 min. A rolling window is then applied to the 10 min output sequence: each window takes its first Kmax saved snapshots as model input and the following Bmax snapshots as the target increments, and is advanced along the time axis by a stride of two output intervals, so that successive samples start from different bed states while remaining only partly overlapping. Using the same counting convention as the other benchmarks, this converts 4,800 raw output intervals into 2,400 input-output samples. For the Waiho River case, eight steady-state scenarios are generated with the same 6:1:1 case-level split. Each Waiho scenario is simulated for 30 h, the duration of the reference flood event, which is long enough for the main erosion corridors and depositional bars to develop on the field-scale mesh, and fields are saved every 10 min; after rolling-window processing, the benchmark produces 1,800 input-output samples. For the Yen case, twenty unsteady hydrograph scenarios are generated: 12 for training, 4 for validation, and 4 for testing. Each Yen scenario is simulated for 5 h, the full length of the experimental hydrograph, so that the rising limb, the peak, and the falling limb are all represented, with a 1 min sampling frequency that resolves the rapid bed adjustment on the rising limb, which is the most demanding phase for the surrogate. In preprocessing, base\_step = 2 and bundle = 2 are used, converting 6,000 raw snapshots into 1,500 model samples. Model performance is evaluated using area-weighted metrics so that nonuniform mesh density does not bias the reported accuracy. Spatial RMSE, domain-mean bed change, cross-sectional bed-elevation change, and area-weighted R\textsuperscript{2} are computed over the same physical domains and output times as the GAIA reference. The validated TELEMAC-2D/GAIA simulations are used to construct the machine-learning dataset for each case, and both the GNO one-step baseline and AH-GNO are trained on the same split.

\begin{figure}[tbp]
\centering
\includegraphics[width=\linewidth,height=0.72\textheight,keepaspectratio]{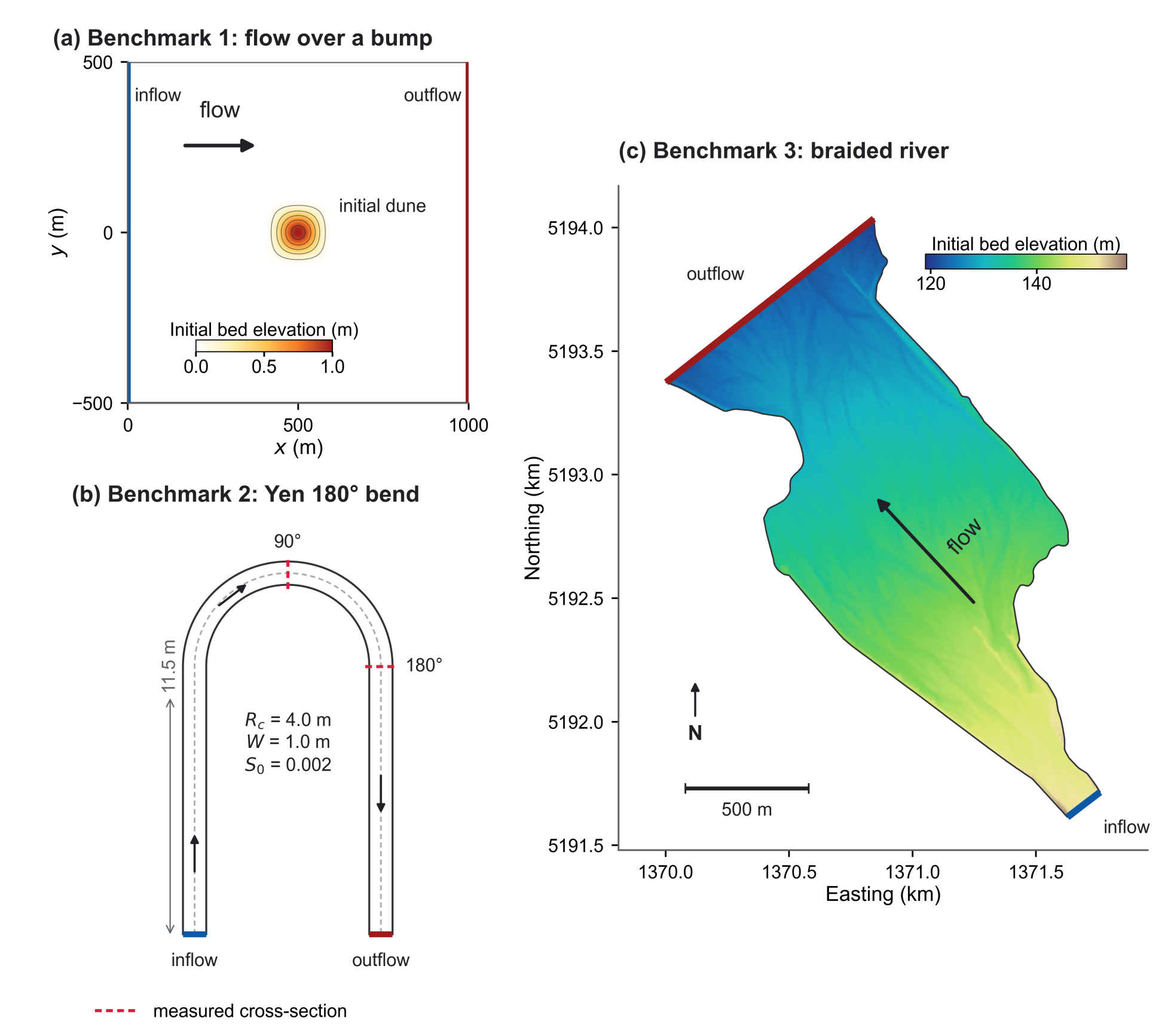}
\caption{The three benchmark cases used to evaluate AH-GNO. (a) Bump: rectangular domain, inflow and outflow boundaries, flow direction, and the initial localized dune. (b) Yen 180\textdegree{} bend: flume geometry with bend radius $R_c$, channel width W and bed slope $S_0$, boundaries, and the measured cross sections at 90\textdegree{} and 180\textdegree{}. (c) Waiho River: field-scale domain, boundaries, flow direction, and initial bed elevation. Colour scales in (a) and (c) give initial bed elevation.}
\label{fig:4}
\end{figure}

\begin{table}[tbp]
\caption{Benchmark data construction and case split for AH-GNO training and evaluation.}
\label{tab:benchmark}
\centering
\small
\begin{tabular}{@{}p{0.17\linewidth}c p{0.27\linewidth}c p{0.14\linewidth}r@{}}
\toprule
Benchmark & Scenarios & Split & Duration & Output & Samples \\
\midrule
Bump & 8 & 6 train / 1 validation / 1 test & 100 h & 10 min & 2,400 \\
Yen 180\textdegree{} bend & 20 & 12 train / 4 validation / 4 test & 5 h & 1 min & 1,500 \\
Waiho River & 8 & 6 train / 1 validation / 1 test & 30 h & 10 min & 1,800 \\
\bottomrule
\end{tabular}\par\vspace{2pt}{\footnotesize\emph{Note:} The source table specifies a stride of two output intervals for Bump and Waiho River.}
\end{table}

\subsection{Benchmark 1: flow over a bump}

The Bump case is a two-dimensional conical-dune test from the GAIA validation suite. The scenario describes the downstream propagation and lateral spreading of a localized sand-dune perturbation under approximately uniform subcritical flow. The computational domain is a 1000 m $\times$ 1000 m rectangle discretized by an unstructured triangular mesh with 2,601 nodes and 5,000 elements. The initial bed is flat except for a local depositional perturbation in the region x = 400-600 m and y = -100-100 m. Sediment transport in GAIA is computed using the Grass total-load formula with AG = 0.00167 m\textsuperscript{2}/s and mg = 3 (Grass, 1981), and only bed-load transport is considered. To evaluate scenario-level robustness, eight steady state discharge-water-level combinations are simulated using the official GAIA setup as the base configuration; detailed inflow and downstream water-level combinations are provided in Table S1 and Figure S1 of the Supporting Information.

Figure 5 compares cumulative bed-elevation change predicted by GAIA, the GNO one-step, and AH-GNO at representative times. The GNO one-step is a non-adaptive baseline that shares the same graph-operator architecture as AH-GNO but writes back only a single predicted bed increment at each coupling call, i.e., the bed is advanced one step at a time with a fixed horizon. AH-GNO adaptively adjusts the effective write-back horizon online according to the evolving bed state during the coupled simulation. The comparison between the GNO one-step and the AH-GNO highlights the effectiveness of adaptive temporal bundling on both enhancing accuracy and efficiency; therefore, the following benchmark cases also follow this assessment protocol.

Results show that the AH-GNO accurately reproduces the main downstream migration path and the progressive lateral spreading of the bedform. The largest deviations occur near the moving bedform margins, especially where bed-elevation gradients are large and the front location is sensitive to small phase errors. The RMSE relative to GAIA increases from 0.0002 m at 1500 min to 0.0007 m at 6000 min, from about 0.2\% to about 0.7\% of the bedform amplitude. Long-term error accumulation is therefore mild. Compared with the GNO one-step, AH-GNO has markedly lower RMSE, showing that adaptive temporal bundling and write-back reduce phase drift and accumulated error during recursive rollout. The lateral-spreading behavior is further evaluated using the 0.1 m isoline at 80 and 100 h (Figure 6). The spreading angle is 25.76\textdegree{} for GAIA and 26.04\textdegree{} for AH-GNO. Thus, AH-GNO reproduces not only the bed-elevation field but also the lateral expansion rate of the bedform structure. Both numerical angles exceed the analytical reference angle of 21.787\textdegree{} (Hudson \& Sweby, 2005; Vriend, 1987), but AH-GNO remains highly consistent with the GAIA reference solution.

\begin{figure}[tbp]
\centering
\includegraphics[width=\linewidth,height=0.72\textheight,keepaspectratio]{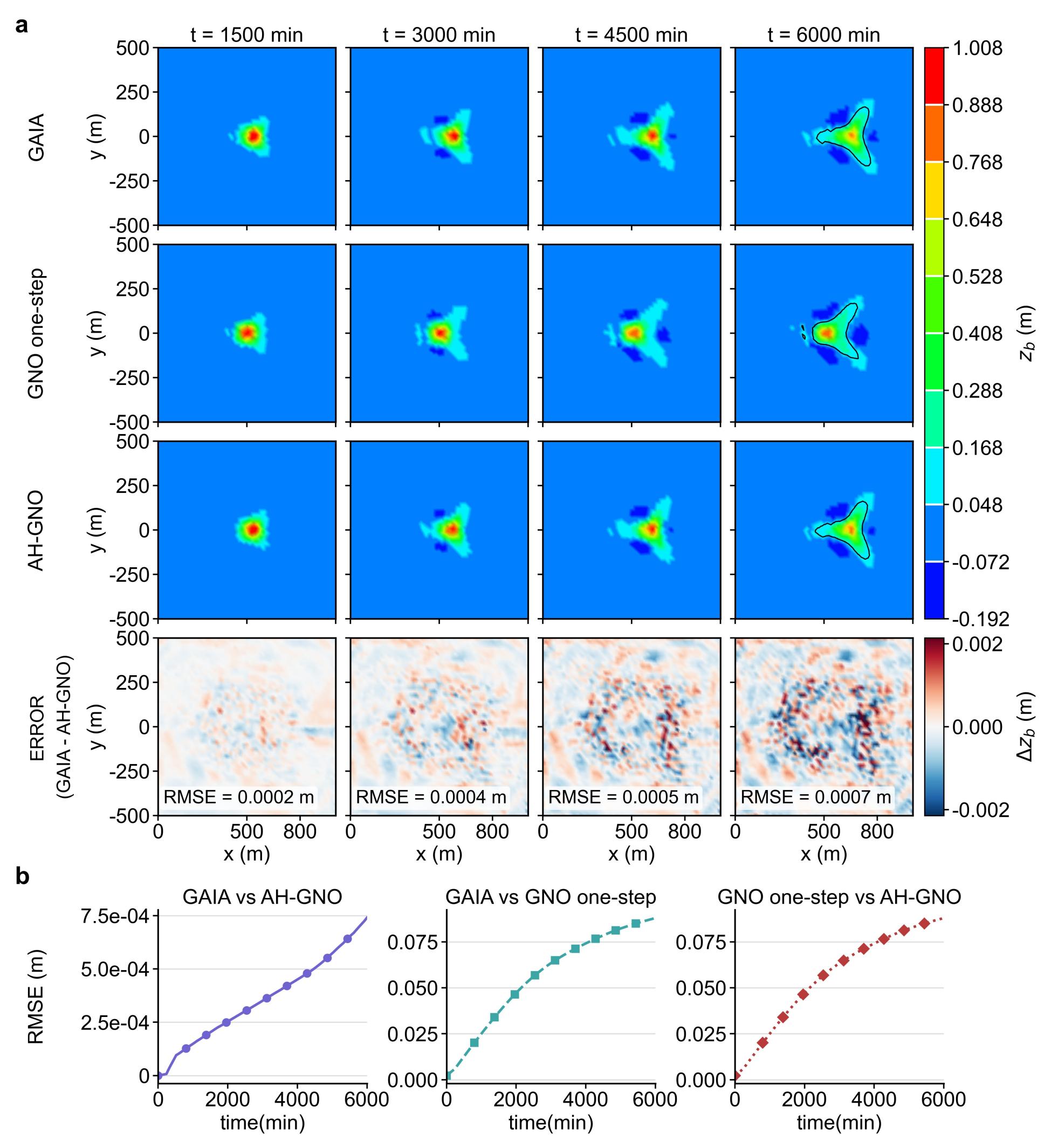}
\caption{Comparison of cumulative bed-elevation change for the Bump benchmark scenario from GAIA, GNO one-step, and AH-GNO, with prediction error relative to GAIA.}
\label{fig:5}
\end{figure}

\begin{figure}[tbp]
\centering
\includegraphics[width=\linewidth,height=0.72\textheight,keepaspectratio]{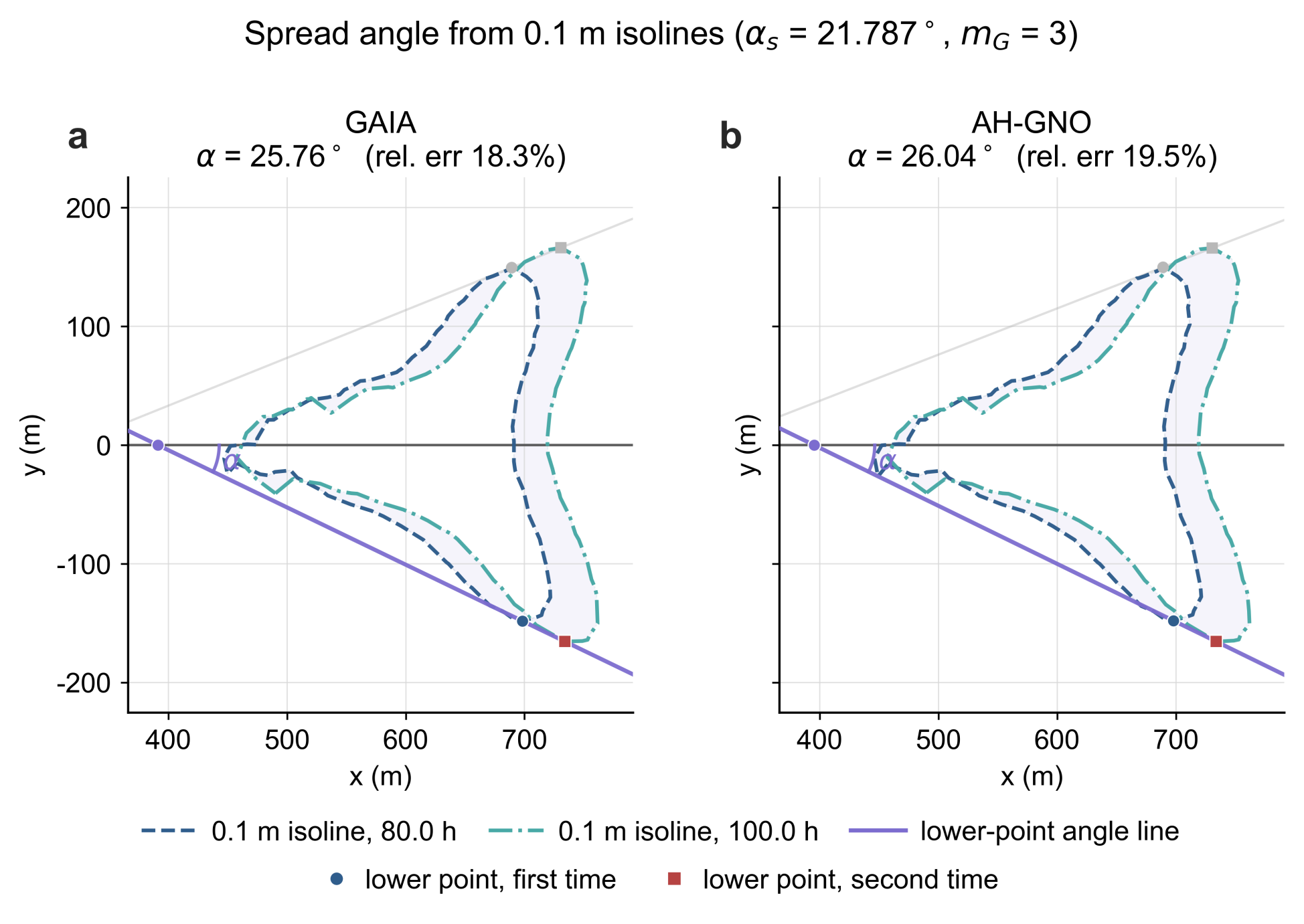}
\caption{Lateral spreading angle for the Bump benchmark extracted from the 0.1-m bed-elevation isoline at 80 h and 100 h for (a) GAIA and (b) AH-GNO methods.}
\label{fig:6}
\end{figure}

\subsection{Benchmark 2: Yen 180\textdegree{} bend}

The Yen case is also from the GAIA validation suite and simulates bed evolution in a 180\textdegree{} bend under unsteady flow. This case is more challenging than the Bump benchmark because bed evolution is affected not only by streamwise sediment transport but also by curvature-induced secondary flow, lateral sediment redistribution, and hydrograph-driven acceleration and deceleration. The case originates from the bend-flume experiments of Yen and Lee \sourcehighlight{(Yen \& Lee, 1995)}. The official Yen configuration is used as the reference physical model and includes a 1.0-m-wide 180\textdegree{} bend, nonuniform sediment, the Engelund-Hansen transport closure, bed-slope effects, and a secondary-flow correction. The unstructured triangular finite-element mesh contains 1,799 nodes and 3,230 triangular elements, with a mean mesh size of approximately 0.20 m and a computational time step of 0.5 s. The full flume geometry, sediment properties, hydrographs, and TELEMAC/GAIA parameters are provided in Table S2 and Figure S2.

Figure 7 compares cumulative bed-elevation change from GAIA, the GNO one-step, and AH-GNO at 30, 50, 90, and 300 min. GAIA shows the expected bend-scale erosion-deposition pattern, with strong erosion along the outer bank and deposition along the inner bank. AH-GNO captures this asymmetric structure at all selected times and maintains agreement with GAIA in the locations of the outer-bank erosion belt and inner-bank deposition zone. In contrast, the GNO one-step shows larger deviations during rapidly varying portions of the hydrograph, indicating that single-step recursive updating is more susceptible to phase and amplitude errors in unsteady bend morphodynamics.

The RMSE evolution in Figure 6 shows that the area-weighted RMSE of AH-GNO relative to GAIA increases briefly during the rising limb of the hydrograph and then stabilizes, remaining overall on the order of 10-4 to 10-3 m. By comparison, errors between the GNO one-step and GAIA are substantially larger and show stronger deviations during rapid adjustment. These results indicate that adaptive historical selection and adaptive write-back can suppress accumulated rollout error while preserving the main morphodynamic features of GAIA.

Figure 8 compares normalized bed-elevation change along the 180\textdegree{} and 90\textdegree{} cross sections against laboratory measurements and GAIA results. At the 180\textdegree{} section, AH-GNO closely matches GAIA over most of the channel width and accurately reproduces the transition from positive bed change on one side of the section to negative bed change on the other. At the 90\textdegree{} section, AH-GNO also remains close to GAIA, although both models underrepresent the large changes observed near the section edges in the experiment. This discrepancy should not be interpreted as a deficiency of AH-GNO itself, because the model is trained on GAIA output rather than on laboratory measurements. The result shows that AH-GNO learns the morphodynamic response represented by the process model, and systematic differences between GAIA and the laboratory observations are inherited by the surrogate.

\begin{figure}[tbp]
\centering
\includegraphics[width=\linewidth,height=0.72\textheight,keepaspectratio]{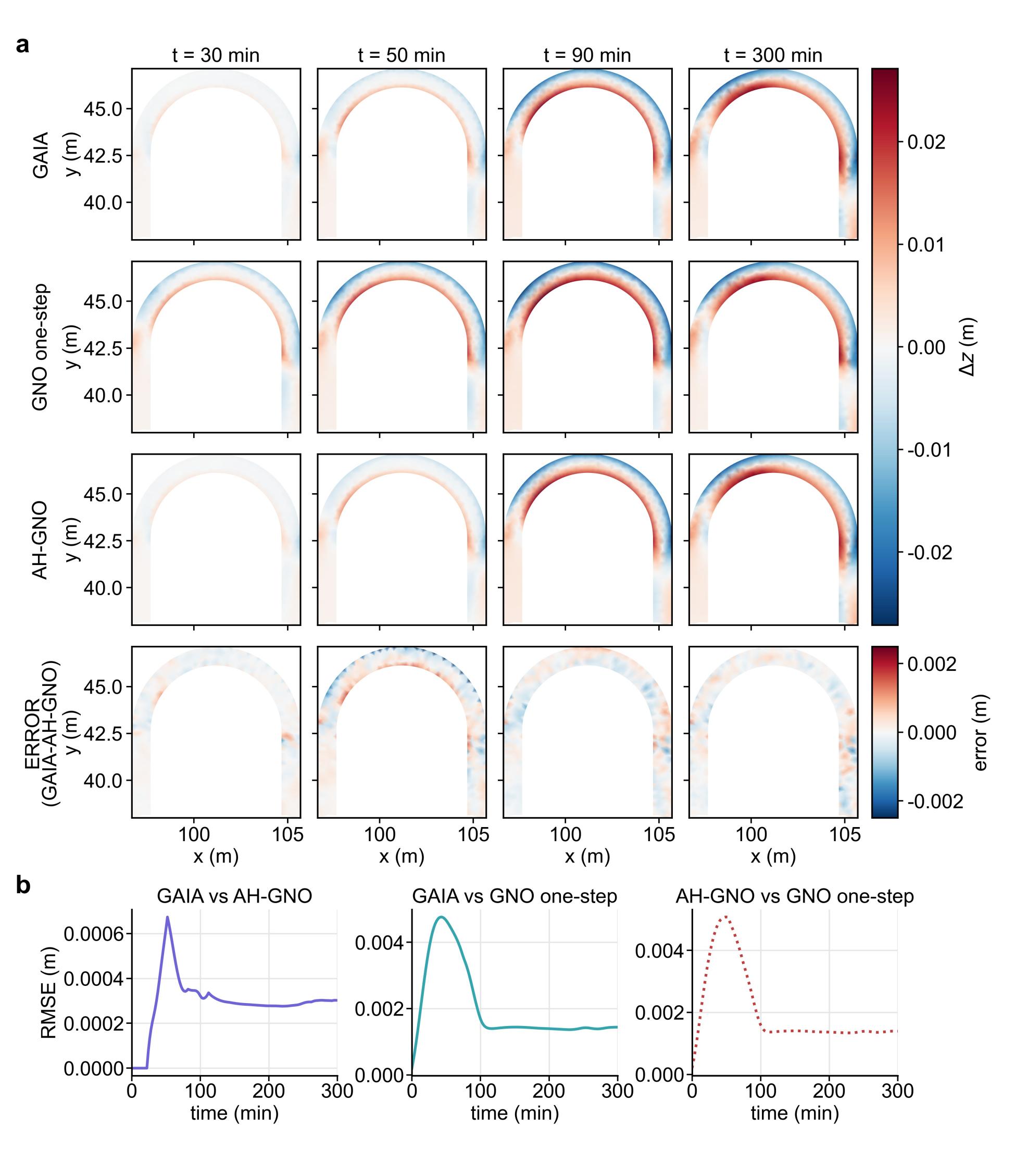}
\caption{Comparison of cumulative bed-elevation change for the Yen 180\textdegree{} bend scenario from GAIA, GNO one-step, and AH-GNO at selected hydrograph times, together with error evolution relative to GAIA.}
\label{fig:7}
\end{figure}

\begin{figure}[tbp]
\centering
\includegraphics[width=\linewidth,height=0.72\textheight,keepaspectratio]{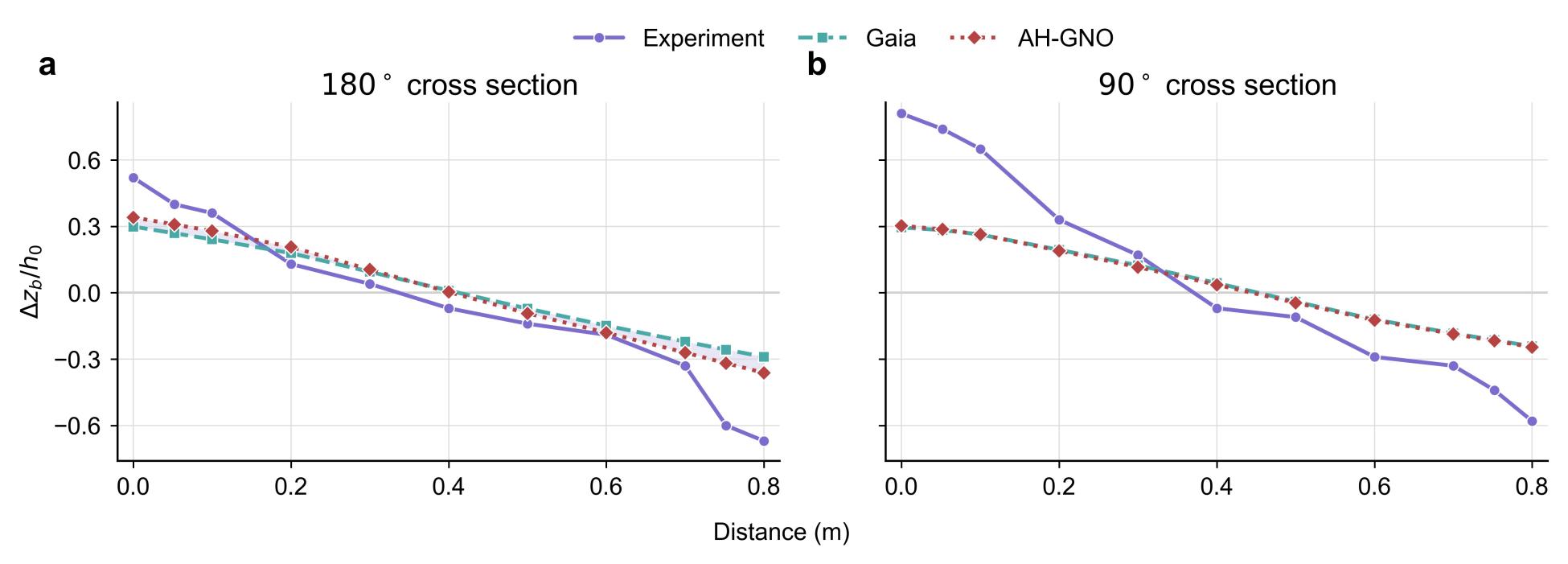}
\caption{Cross-sectional bed-elevation change in the Yen bend benchmark. Panels a and b correspond to the 180\textdegree{} and 90\textdegree{} sections and compare laboratory observations, GAIA, and AH-GNO.}
\label{fig:8}
\end{figure}

\subsection{Benchmark 3: braided river}

The third case is a field-scale braided river application for the Waiho River in New Zealand. Compared with the two official benchmarks, the Waiho River has more complex topographic boundaries, braided-channel connectivity, alternating local pools and bars, and more pronounced wetting-drying transitions. The Waiho River is located on the west coast of New Zealand\textquotesingle s South Island and drains the western slope of the Southern New Zeeland. The river receives a high sediment supply from the Franz Josef Glacier and the Callery catchment, which promotes long-term aggradation in the river. However, the numerical experiments in this study use a controlled setting without an external sediment supply. The purpose is to isolate and test the ability of the bed-update operator to learn redistribution of existing bed material, while avoiding confounding effects from uncertain field sediment inputs, transport-rate curves, and boundary sediment supply. The Waiho case should therefore be interpreted as a controlled morphodynamic benchmark under field-scale geometry and hydrodynamics, not as a complete reproduction of the river\textquotesingle s natural morphological evolution.

The Waiho River reference simulation is built using TELEMAC-2D and GAIA modules. Hydrodynamic variables and bed elevation are defined on the native unstructured triangular mesh, which contains 19,335 nodes and 37,976 triangular elements and covers approximately 1.64 km\textsuperscript{2}. The hydrodynamic time step is 5 s, and the physical simulation duration is 30 h. The GAIA configuration includes four noncohesive sediment classes with diameters of 0.0015, 0.0080, 0.0320, and 0.0900 m and a density of 2650 kg/m\textsuperscript{3}. Only bed-load transport is considered, using a Wilcock-Crowe-type setting. Eight discharge-water-level combinations are generated from the Waiho setup: six for training, one for validation, and one for testing. Detailed inflow hydrographs, downstream water-level conditions, and model parameters are provided in Table S3 and Figure S3.

Figure 9 compares cumulative bed-elevation change from GAIA, the GNO one-step, and AH-GNO at 5, 12, and 30 h. GAIA shows erosion and deposition distributed across multiple active channels, with local bed lowering in the upstream contraction area, the main erosion corridor, and downstream active-channel zones. The GNO one-step captures the approximate spatial extent of active channels, but differs more strongly in local change magnitude and domain-mean bed change. AH-GNO more accurately reproduces the main erosion corridors and adjacent depositional belts, and the spatial pattern at the final 30 h stage remains highly consistent with GAIA.

In terms of domain-mean bed change, GAIA gives mean increments of -0.010, -0.017, and -0.028 m at 5, 12, and 30 h, respectively. The GNO one-step gives -0.014, -0.015, and -0.023 m, whereas AH-GNO yields -0.008, -0.018, and -0.027 m. AH-GNO preserves the sign of net erosion and the increasing trend over time, with a final mean value close to GAIA (within about 0.001 m), although the profile remains somewhat smoother in highly localized scour and deposition zones. The GNO one-step more clearly underestimates net erosion at later times. The residual smoothing of local extreme erosion and deposition is not unexpected for neural surrogate models: neural networks have a documented spectral bias toward lower-frequency functions \sourcehighlight{(Rahaman et al., 2019)}, and recent neural-operator studies report that high-frequency or sharp-gradient content can be attenuated unless special high-frequency corrections are introduced \sourcehighlight{(Khodakarami et al., 2025)}.

\begin{figure}[tbp]
\centering
\includegraphics[width=\linewidth,height=0.72\textheight,keepaspectratio]{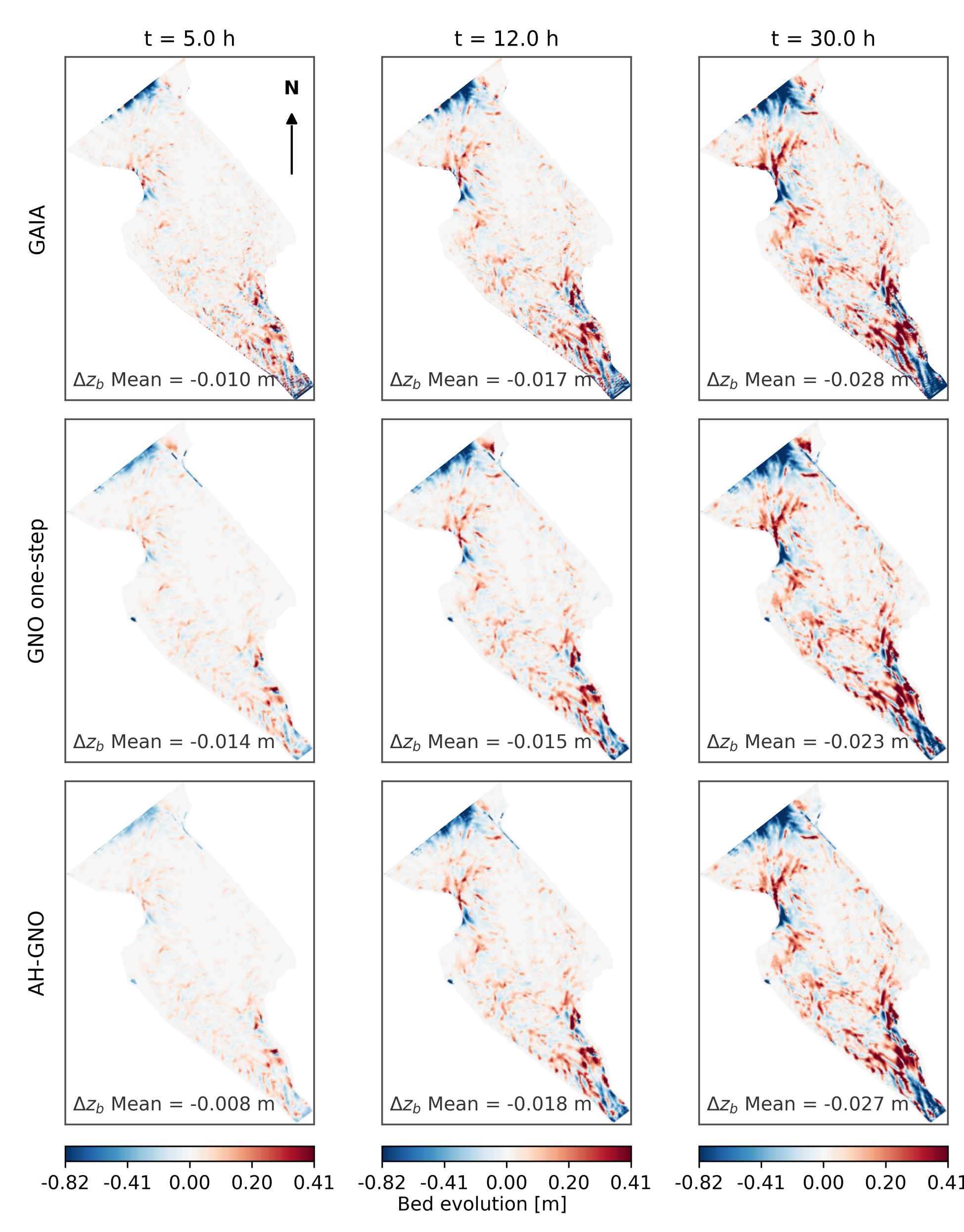}
\caption{Waiho River field-scale benchmark comparison of cumulative bed-elevation change from GAIA, GNO one-step, and AH-GNO at 5, 12, and 30 h.}
\label{fig:9}
\end{figure}

To further assess whether map-scale consistency is retained in one-dimensional topographic diagnostics, Figure 10 compares GAIA and AH-GNO along three representative transects in the Waiho River domain. Section A crosses an active downstream channel, where both models reproduce the main depositional peak and subsequent small bed undulations. Section B intersects a local erosion feature; AH-GNO captures the location and direction of the sharp bed lowering, but gives a smoother profile near the deepest local trough. Section C covers an upstream part of the study domain, where the largest difference occurs in the first 100 m of the transect, while the remaining profile is close to GAIA. These transect-based analyses show that the neural operator preserves the main cross-sectional morphodynamic features while smoothing some localized extremes.

\begin{figure}[tbp]
\centering
\includegraphics[width=\linewidth,height=0.72\textheight,keepaspectratio]{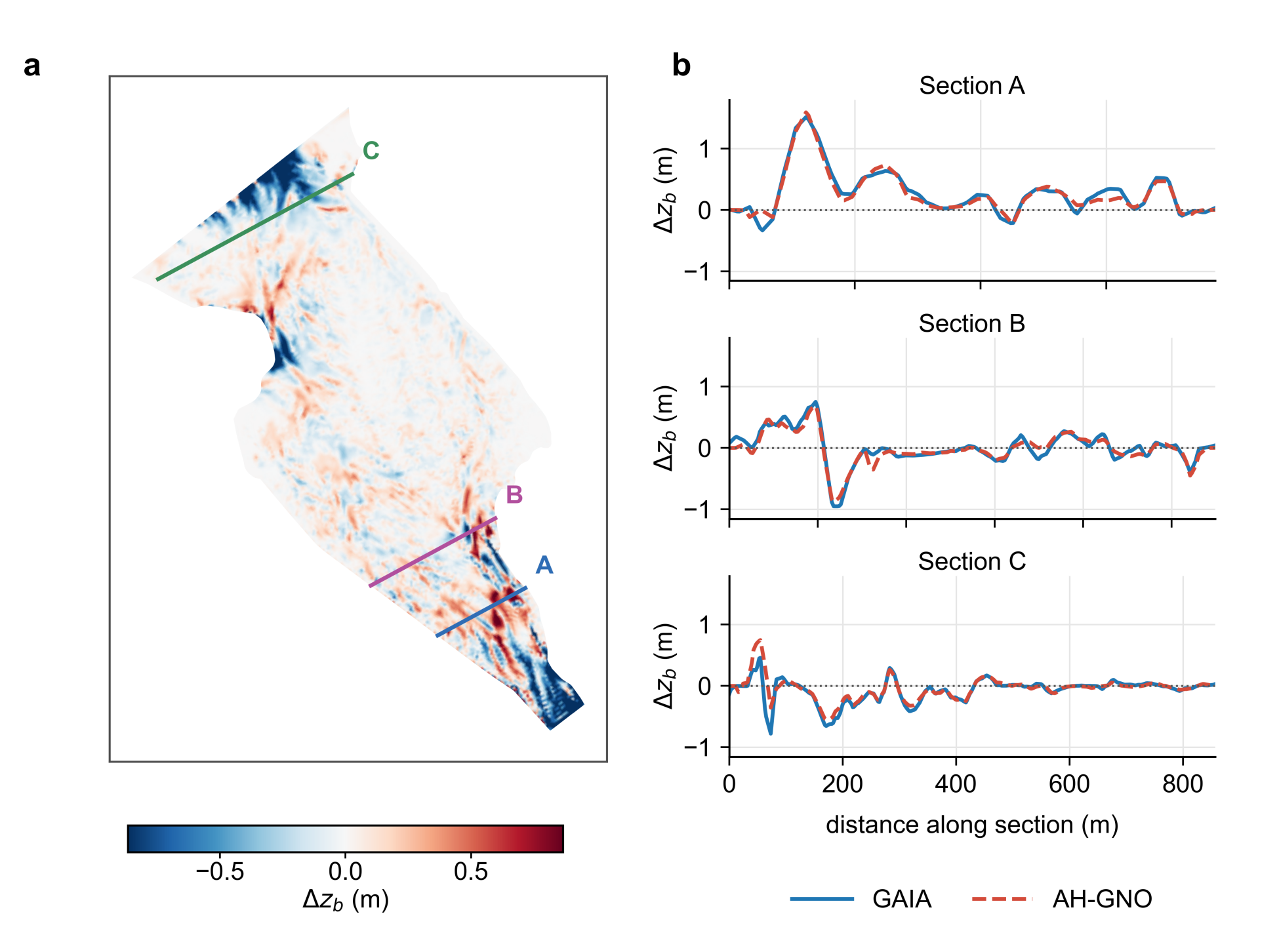}
\caption{Waiho River cross-sectional diagnostic of cumulative bed-elevation change. (a) Locations of sections A-C over the cumulative bed-change field. (b) Cumulative bed-elevation change along the three sections from GAIA and AH-GNO. AH-GNO captures the main depositional and erosional peaks along the selected transects, with smoother responses at localized extremes.}
\label{fig:10}
\end{figure}

\subsection{Runtime performance in coupled simulations}

To evaluate computational efficiency, three bed-evolution approaches are compared using their actual wall-clock runtimes. For each benchmark case, all three approaches use the same computational mesh, physical simulation duration, hydrodynamic time step, and output frequency, while the processor allocation is varied explicitly: GAIA is evaluated using both 4- and 20-core CPU configurations, whereas the neural surrogate models are evaluated on an NVIDIA GeForce RTX 4090 GPU. For the neural surrogate models, the reported runtime includes model inference, input normalization, output denormalization, accumulation of bed-elevation increments, and variable exchange with TELEMAC-2D during online coupling, but excludes offline GAIA data generation and model training. The speedups reported here should therefore be interpreted as post-training, deployment-specific online-inference speedups relative to the CPU-based GAIA bed-update workflow for the test scenarios, not as speedups for replacing the entire TELEMAC-MASCARET modeling process or as hardware-independent acceleration factors.

Figure 11 shows that replacing the explicit GAIA bed-update calculation with a learned graph operator (GNO or AH-GNO) substantially reduces computational cost under both GAIA CPU allocations. In the Bump case, runtimes are 228.1 and 123 s for GAIA using 4 and 20 CPU cores, respectively, compared with 15.31 s for the GNO one-step and 0.66 s for AH-GNO. Relative to the 4-core GAIA baseline, the two neural approaches achieve approximate speedups of 15 and 346 times; even relative to the 20-core GAIA result, their speedups remain approximately 8 and 186 times. In the Yen bend case, the corresponding runtimes are 12.0 and 8.98 s for the 4- and 20-core GAIA runs, 0.74 s for the GNO one-step, and 0.08 s for AH-GNO. These values correspond to approximate GNO and AH-GNO speedups of 16 and 150 times relative to 4-core GAIA, and 12 and 112 times relative to 20-core GAIA. In the Waiho River case, the respective runtimes are 1204 and 352 s for the two GAIA configurations, 33.41 s for the GNO one-step, and 3.34 s for AH-GNO. The resulting speedups are approximately 36 and 360 times relative to 4-core GAIA, and 10.5 and 105 times relative to 20-core GAIA.

Increasing the GAIA allocation from 4 to 20 CPU cores reduces runtime by factors of approximately 1.9, 1.3, and 3.4 for the Bump, Yen, and Waiho cases, respectively. The stronger parallel benefit in the Waiho case is consistent with the substantially larger mesh and computational workload, whereas the smaller Bump and Yen meshes provide less work per processor and are therefore more susceptible to parallelization overhead. Importantly, the GNO one-step remains approximately 8-12 times faster, and AH-GNO remains approximately 105-186 times faster than the corresponding 20-core GAIA runs. Thus, although the absolute speed-up depends on processor type, allocation, implementation, and problem size, the additional 20-core results confirm that the substantial computational advantage of the GNO-based approach is not restricted to the original 4-core GAIA baseline.

The comparison between the GNO one-step and AH-GNO further clarifies the computational role of adaptive temporal bundling. Both neural models replace the sediment transport calculation and the Exner-type bed update in GAIA with a graph operator; therefore, the GNO one-step already reduces the cost of an individual bed update. However, the GNO one-step writes back only one predicted bed increment at each coupling call and thus still requires frequent calls to the neural bed-update module. AH-GNO, by contrast, predicts a bundle of future increments at each inference step and estimates how many of them can be reliably written back before the next hydrodynamic feedback is needed. The acceleration of AH-GNO therefore arises not only from faster individual bed-update computation, but also from a reduction in the number of bed-update calls. As the GNO one-step and AH-GNO are evaluated on the same GPU platform, their direct runtime comparison is less affected by cross-device differences. AH-GNO is approximately 23.2, 9.3, and 10.0 times faster than the GNO one-step in the Bump, Yen, and Waiho cases, respectively, providing more direct evidence of the computational benefit of adaptive temporal bundling.

This reduction in call frequency is strongly state-dependent. The mean effective write-back lengths in the Bump, Yen, and Waiho River cases are 9.8, 10.6, and 10.1 steps, respectively. Shorter effective write-back lengths typically occur during rapid bed adjustment, secondary-flow-driven lateral transport in bends, enhanced local scour, or pronounced wetting-drying transitions. Longer effective write-back lengths correspond to periods of slower bed change and smaller prediction uncertainty. The adaptive temporal-bundling mechanism therefore establishes a state-dependent compromise between computational efficiency and rollout stability: the approach reduces unnecessary bed-update calls during slowly evolving periods while avoiding excessive extrapolation during rapid morphodynamic change.

Online compute cost and offline training cost should be interpreted separately. AH-GNO contains approximately 0.346 million trainable parameters. The offline training times for the Bump, Yen, and Waiho River cases are 7.90, 3.12, and 18.08 h, respectively. Although training requires supervised data from GAIA, the trained model can be reused across multiple similar scenarios. The main advantage of AH-GNO, therefore, lies in applications requiring repeated morphodynamic simulations, such as scenario ensembles, sensitivity analysis, parameter screening, and engineering design comparison. In such applications, the initial data generation and training costs can be moderated over many online inference runs.

\begin{figure}[tbp]
\centering
\includegraphics[width=\linewidth,height=0.72\textheight,keepaspectratio]{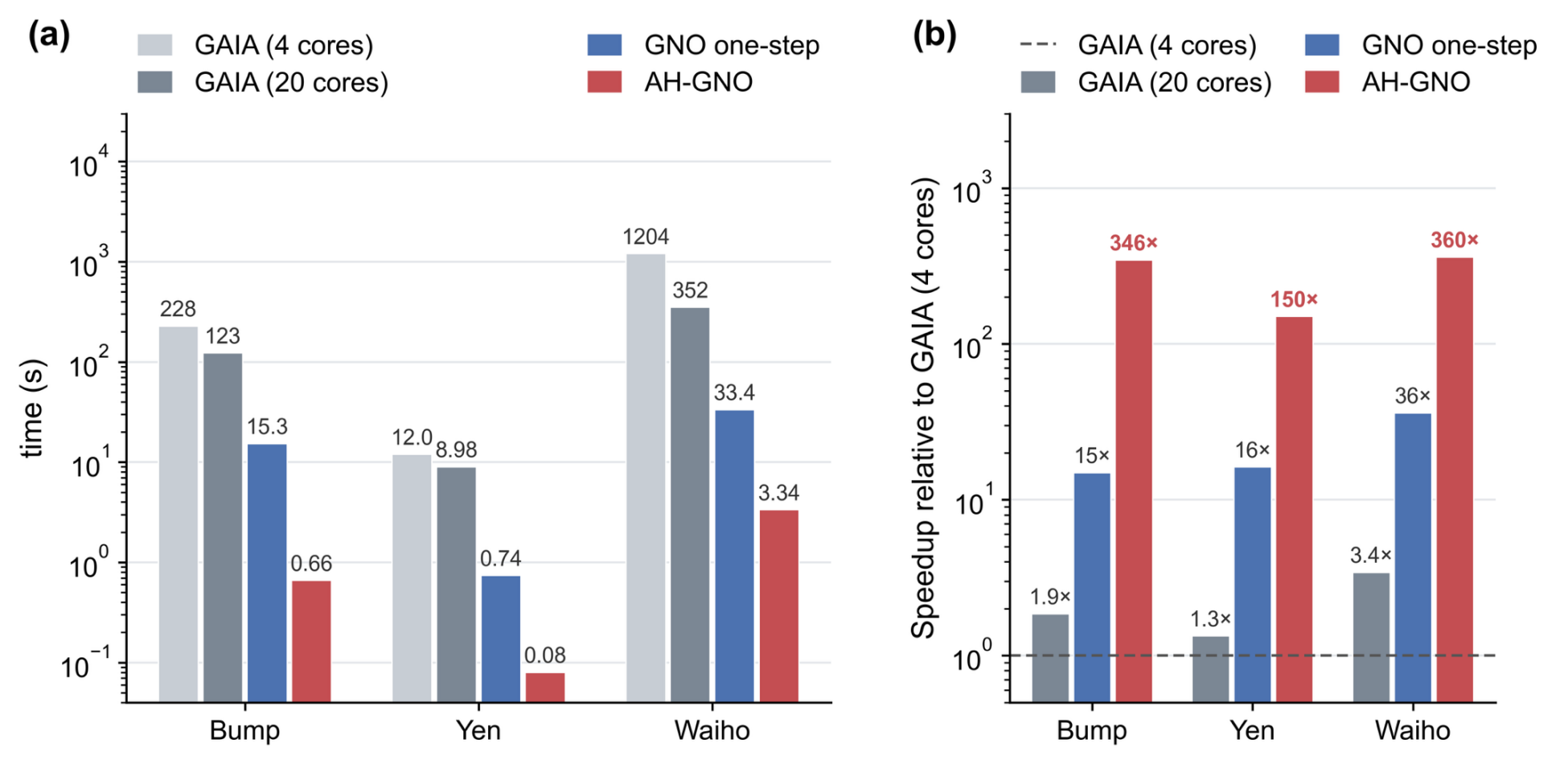}
\caption{Runtime performance: (a) wall-clock runtime comparison of GAIA using 4 and 20 CPU cores, GNO one-step, and AH-GNO for the three benchmark cases; (b) computational speedup relative to GAIA running on the 4-core CPU configuration.}
\label{fig:11}
\end{figure}

These results demonstrate the computational feasibility of replacing the explicit GAIA bed-update step with a learned graph operator. The persistence of substantial acceleration when the GAIA allocation is increased from 4 to 20 CPU cores strengthens this conclusion, while the comparison also shows that runtime ratios should be reported together with their hardware configurations rather than interpreted as universal constants. This acceleration is particularly relevant to morphodynamic applications, where engineering practice commonly requires many simulations for hydrograph scenarios, sensitivity tests, calibration, or design comparison. As TELEMAC-2D remains online, the system does not need to be converted into an open-loop neural time-series model; AH-GNO accelerates the morphodynamic update while retaining hydrodynamic feedback from the process solver.

\subsection{Discretization consistency}

Meshes used in morphodynamic applications are frequently refined locally near structures, riverbanks, and active channels, or rebuilt from updated bathymetric surveys, and may differ from the mesh used for surrogate training. We define discretization consistency as the ability of a trained AH-GNO to reproduce the bed-elevation change field on meshes of different resolution and node distribution without retraining or interpolation. This property follows from the operator formulation, which maps between spatial fields rather than fixed-length nodal vectors. Four meshes of the Yen bend were tested: a coarse 0.3 m mesh, the original mesh, a fine 0.1 m mesh, and an irregular mesh. TELEMAC-2D/GAIA reference simulations were run on each mesh with identical physical settings, sediment parameters, and boundary forcing. The AH-GNO trained on the original mesh was applied to all four meshes without modification. The GNO one-step baseline, which shares the area-weighted graph integral operator governing mesh response, was not tested separately. The test does not demonstrate strict mesh independence; predictions remain sensitive to node distribution, search radius, and local mesh irregularity. Figure 12 shows the evolution of area-weighted R\textsuperscript{2} for representative hydrograph scenarios on the four meshes.

In most cases, area-weighted R\textsuperscript{2} increases rapidly during the initial stage of bed change and remains high after approximately 60-120 min. Cases 01, 03, 07, 11, 14, 15, and 19 (see Table S2) approach R\textsuperscript{2} values near 0.99 on all four meshes during the later development stage. This value indicates that area-weighted graph integration and physically distance-based neighborhood aggregation substantially reduce sensitivity to node density and mesh layout, allowing the same learned parameters to represent cumulative bed-change fields across different discretizations.

Some hydrographs are more sensitive to discretization. Case 09 is the most challenging example: the irregular bend mesh shows a late-stage decrease in R\textsuperscript{2}, whereas the other meshes remain relatively accurate. Cases 08, 12, and 18 also yield slightly lower values than the simplest hydrographs. These decreases suggest that local erosion-deposition patterns can interact with mesh layout, especially when active bed change occupies only a limited part of the channel or when sharp morphodynamic fronts are represented by different local node distributions

Figure 13 summarizes the multi-mesh results. Averaged over all scenarios, the coarse 0.3-m mesh gives the highest mean R\textsuperscript{2}, followed by the original Yen mesh, the fine 0.1-m mesh, and the irregular bend mesh; the mean values in the heat map are approximately 0.993, 0.991, 0.985, and 0.981, respectively. The lower R\textsuperscript{2} on the irregular mesh is expected because the case differs most strongly from the training discretization, but the mean value nonetheless remains above 0.98. Across all mesh-scenario combinations, mean R\textsuperscript{2} ranges from approximately 0.951 to 0.998, with an overall mean of about 0.987.

The spread in Figure 13c across the different meshes further indicates that Case 09 has the strongest discretization sensitivity, followed by Case 12, whereas Cases 03 and 11 show very small mesh-to-mesh differences. These results support the practical cross-mesh applicability of AH-GNO on unstructured morphodynamic meshes, while also showing that uncertainty monitoring remains necessary when the hydrological response is highly localized or the target mesh differs substantially from the training discretization.

\begin{figure}[tbp]
\centering
\includegraphics[width=\linewidth,height=0.72\textheight,keepaspectratio]{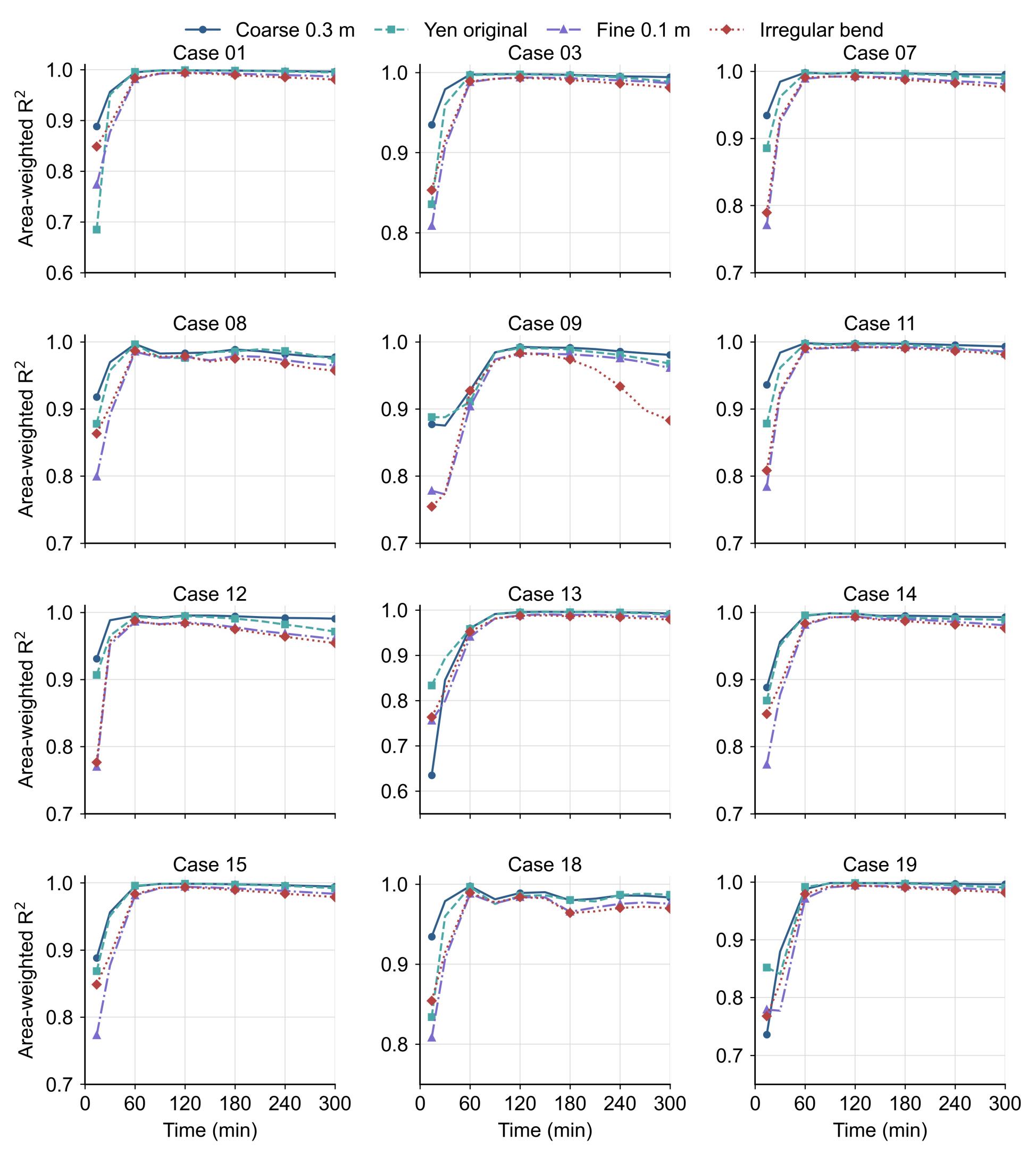}
\caption{Area-weighted R\textsuperscript{2} evolution for AH-GNO on different Yen bend meshes and hydrograph scenarios.}
\label{fig:12}
\end{figure}

\begin{figure}[tbp]
\centering
\includegraphics[width=\linewidth,height=0.72\textheight,keepaspectratio]{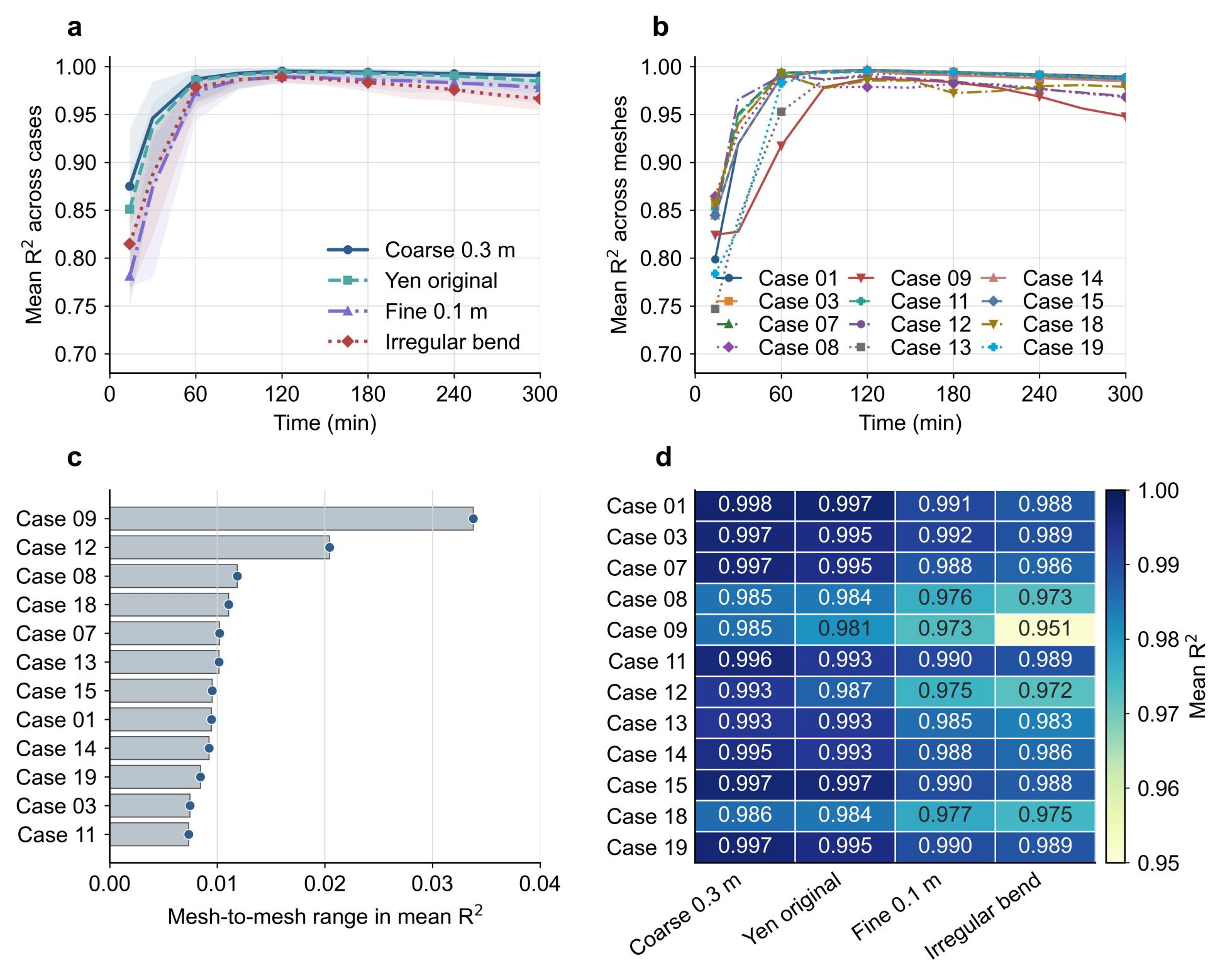}
\caption{Cross-mesh discretization consistency analysis. a Mean R\textsuperscript{2} over scenarios for each mesh, with shading indicating scenario variability; b mean R\textsuperscript{2} over meshes for each scenario; c mesh-to-mesh R\textsuperscript{2} spread for each scenario; d heat map of mean R\textsuperscript{2} for each scenario-mesh combination.}
\label{fig:13}
\end{figure}

\section{Discussion}

\subsection{Limitations}

The primary limitation of this study is that explicit sediment transport state variables are not currently included in the AH-GNO input. The present model uses water depth, the two depth-averaged velocity components, and bed elevation to predict increments in bed elevation. However, the approach does not directly use morphodynamic variables such as bed shear stress, Shields stress, sediment transport rate, sediment concentration, active layer thickness, grain size fractions, spatially varying D50, or upstream sediment supply. AH-GNO should therefore be interpreted as a lightweight operator for bed updating, trained under the sediment closure and sediment state assumptions of the reference GAIA simulations, rather than as a complete replacement for a mechanistic sediment transport module.

This simplification does not invalidate the present benchmarks. The Bump and Yen cases are bed-load-dominated GAIA validation problems, and the Waiho experiment is designed as a controlled field-scale redistribution test. Within each case, the sediment parameters, transport closure, and boundary settings are fixed across scenarios. Under these conditions, local water depth, velocity, and existing bed morphology contain much of the information needed to learn the dominant bed response produced by GAIA, while avoiding additional exchange of sediment variables during online coupling. The results, therefore, show that a lightweight operator for bed increments can be useful under idealized or controlled sediment settings. The approach also provides a basis for future work extending to use of operators that also include sediment state information.

The absence of sediment-state variables from the input also limits transfer to problems in which the bed sediment state changes strongly. In its current form, AH-GNO does not represent active layer coarsening or fining, changes in grain-size composition, hiding and exposure effects beyond those implicitly contained in the training data, strong suspended load transport, hyperconcentrated flow, substantial external sediment supply, or sediment replenishment controlled by the boundary. In such settings, hydrodynamics and bed elevation alone do not uniquely determine the future bed update. Missing sediment state information may lead to underestimated erosion or deposition, delayed bed adjustment, inadequate response of individual grain size classes, or incorrect recovery after changes in sediment supply. The present results should therefore be interpreted as evidence for accelerating bedload-dominated morphodynamic simulations under controlled sediment conditions, rather than as evidence of general applicability to all sediment regimes.

The Waiho River case requires particular caution. The natural Waiho River receives substantial sediment from glacier-fed and mountain catchments, whereas the numerical experiment used here intentionally excludes external sediment replenishment. This upstream supply cutoff was adopted because field sediment inputs, transport rate relations, and boundary sediment supply are highly uncertain, and their inclusion would make it difficult to isolate the error introduced by replacing the GAIA bed update with AH-GNO. The Waiho comparison is therefore most appropriate for evaluating whether AH-GNO can reproduce the spatial distribution and relative intensity of erosion and deposition generated by GAIA under the same controlled setup. The simulations should not be interpreted as quantitatively complete predictions of the natural aggradation or degradation budget of the Waiho River.

GAIA is therefore used as a process-based numerical reference rather than as observational truth. The relevance to this study arises from the explicit coupling inTELEMAC-2D/GAIA of shallow water hydrodynamics, bed load transport, and bed updating on the same realistic river geometry, unstructured mesh, grain size setting, and boundary conditions used by AH-GNO. As both models are evaluated under identical numerical and hydraulic settings, the reported error mainly measures the substitution error introduced by replacing the GAIA bed update calculation with AH-GNO. However, any systematic bias of GAIA relative to field observations, as well as any bias introduced by omitting external sediment supply, will also be inherited by the surrogate comparison.

A second limitation is that the model directly predicts increments in bed elevation and does not explicitly compute sediment flux or flux divergence. It therefore does not automatically satisfy local or global mass conservation as imposed by the Exner equation. The area-weighted loss and area-weighted metrics make the objective closer to a physically integrated error in bed change, but area weighting cannot replace a conservation law. During long rollouts, accumulated erosion and deposition volume, domain mean bed change, and mass errors near boundaries should be monitored, especially where erosion and deposition are intermittent and spatially concentrated.

Online coupling reduces the accumulation of bed-elevation error, but does not remove the cumulative effects entirely. As TELEMAC-2D remains in the computational loop, water depth and velocity are recomputed after each AH-GNO bed update and can respond to the updated topography. This process distinguishes the present framework from an open-loop neural time series model and helps retain hydrodynamic feedback. Nevertheless, if AH-GNO produces a systematic bias during a certain period or in a certain region, TELEMAC-2D will continue the simulation over a biased bed, and secondary errors may develop. Larger errors are expected near shallow areas, inflow and outflow boundaries, wetting and drying zones, fixed boundary regions, and sharply localized active bed features.

The Yen experiment on multiple meshes should be interpreted as evidence of practical mesh robustness rather than absolutely strict mathematical mesh independence. Ordinary graph neural networks learn message passing on a specific graph structure and may be sensitive to fixed adjacency and node distribution. AH-GNO instead approximates a continuous spatial integral operator, discretized by graph-based neighborhood aggregation. Physical coordinates, relative position kernels, and area weights allow the same parameters to operate across different node counts and mesh densities. However, error control and uncertainty assessment are still necessary when active bed change is highly localized or when the target mesh differs substantially from the training discretization.

\subsection{Future Work}

Future work should first include a richer description of sediment state and transport capacity in the model input. Bed shear stress, the dimensionless Shields parameter, active layer thickness, volumetric fractions of individual grain size classes, spatially varying D\textsubscript{50}, near-bed sediment concentration, or equivalent transport capacity could be added as features. These additions would move the model from a surrogate for bed increments toward a morphodynamic component that retains memory of the evolving bed sediment state. For field reaches where external sediment supply is important, upstream input, lateral supply, and empirical or process-based transport relations should also be included in the training design, so that the model does not learn only local redistribution under no supply conditions.

Physical constraints and error control should also be strengthened. The present model does not explicitly satisfy sediment mass conservation in the Exner equation, so future training or inference could include constraints on integrated bed change, penalties on domain scale volume error, local smoothness constraints, and consistency checks near sediment flux boundaries. Another direction would be to let the model jointly predict increments in bed elevation and an equivalent sediment flux divergence, which would connect the neural output more directly to the conservation law. For online coupling, sediment volume error monitors, uncertainty thresholds, or local anomaly detectors could be used. When the accumulated volume error, local bed change, or model uncertainty exceeds a prescribed range, the effective write-back length could be shortened automatically, or the system could temporarily return to GAIA for correction.

The adaptive temporal control also requires further development. The K/B mechanism used here adjusts the effective history length and write-back step according to the current state, but the policy is learned from data and relies on a fixed numerical tolerance. Future studies could supervise it with physical error indicators. This type of adaptive coupling, based on both state and error, would be more suitable for unsteady floods and real river simulations than a fixed coupling interval.

Model generalization should be examined using a broader set of cases. The present study includes the idealized Bump case, the experimental scale Yen bend, and the field scale Waiho River, but these cases do not cover the full range of boundary conditions, channel patterns, sediment mixtures, roughness settings, and flood processes found in natural rivers. Future work should include systematic tests on straight channels, meandering channels, braided rivers, alluvial fans, and engineered reaches. Sample libraries should cover different flood peaks, rising and falling durations, grain size mixtures, initial bed forms, and sediment supply conditions at the boundaries. For reaches outside the training distribution, active learning, a small number of GAIA correction samples, or transfer learning could be used for rapid retraining or fine-tuning.

More efficient coupling strategies are another important direction. This study keeps TELEMAC-2D as the hydrodynamic solver, so the feedback between flow and bed topography remains physically consistent, but the total runtime is still partly controlled by the hydrodynamic calculation. Future work could develop a coupled surrogate system in which one model predicts water depth and velocity, and another predicts bed increments. This approach may further improve the efficiency of large scenario ensembles, but it would also make stability and error control more difficult. If both the hydrodynamic and morphodynamic surrogates contain bias, the errors may amplify through their interaction. Such systems would therefore require stronger conservation constraints, uncertainty estimation, correction by the process model, and fallback mechanisms when the surrogate predictions become unreliable. Overall, the value of AH-GNO is not that it fully replaces traditional morphodynamic models, but instead that this approach provides a fast operator for bed updating that can be embedded in process model loops and applied on unstructured meshes.

\section{Conclusions}

This study presents AH-GNO, an Adaptive-Horizon Graph Neural Operator for online coupled morphodynamic simulation with TELEMAC-2D. The method represents the GAIA bed update as a bed-elevation-increment operator on unstructured meshes and combines area-weighted graph integration, adaptive historical selection, and adaptive forecast-horizon estimation. The operator works on the native computational mesh. At each coupling call it reads the effective history length and the write-back horizon from the current flow-bed state, instead of using values fixed before the run. In this way, the computational cost of bed updating is substantially reduced while the closed-loop hydrodynamic feedback from TELEMAC-2D is retained.

The Bump, Yen, and Waiho River cases show that AH-GNO preserves the main bed-evolution patterns across different levels of complexity, from idealized bedform migration to unsteady bend erosion-deposition and field-scale braided-river applications. Domain-mean bed change and the position and sign of the main erosion and deposition features follow GAIA closely, while sharp scour troughs and depositional margins are smoothed in magnitude. The same learned parameters hold on meshes absent from training, where mean area-weighted R\textsuperscript{2} across the four Yen meshes falls between 0.981 and 0.993. For the tested post-training bed-update workflow, AH-GNO achieved speedups of approximately 346, 150, and 360 times relative to the 4-core GAIA configuration for the Bump, Yen, and Waiho cases, respectively. When GAIA was executed using 20 CPU cores, the corresponding AH-GNO speedups remained approximately 186, 112, and 105 times. Of these margins, a factor of 9 to 23 comes from the adaptive horizon rather than from the graph operator alone, since write-back lengths of about ten steps cut the number of bed-update calls. The end-to-end coupled simulation acceleration also depends on the remaining hydrodynamic computational cost. Margins of this size matter most in work that repeats one reach many times over, such as calibration, sensitivity analysis, flood-scenario ensembles, and comparison of engineering layouts.

These results support the use of learned graph operators as local acceleration components within process-based morphodynamic models rather than as replacements for the complete physical system. TELEMAC-2D stays in the loop and recomputes depth and velocity over every predicted bed, so the surrogate never extrapolates the coupled system on its own. The current method remains limited by training-data coverage, sediment-process representation, boundary treatment, and long-term conservation error. The model predicts bed increments without an explicit sediment flux, and Exner mass conservation is not imposed. The evidence presented here covers bed-load-dominated conditions under a fixed transport closure. With future incorporation of physical constraints, online uncertainty monitoring, and periodic process-model correction, AH-GNO may become a reliable coupled tool for multi-scenario morphodynamic simulation and rapid engineering-design screening.

\section*{References}
\begingroup
\small
\noindent\hangindent=1.6em\hangafter=1 Abid Almubaidin, M. A., Latif, S. D., Balan, K., Ahmed, A. N., \& El-Shafie, A. (2023). Enhancing sediment transport predictions through machine learning-based multi-scenario regression models. \emph{Results in Engineering}, \emph{20}, 101585. \url{https://doi.org/10.1016/j.rineng.2023.101585}\par\smallskip
\noindent\hangindent=1.6em\hangafter=1 Afan, H. A., El-shafie, A., Mohtar, W. H. M. W., \& Yaseen, Z. M. (2016). Past, present and prospect of an artificial intelligence (AI) based model for sediment transport prediction. \emph{Journal of Hydrology}, \emph{541}, 902--913. \url{https://doi.org/10.1016/j.jhydrol.2016.07.048}\par\smallskip
\noindent\hangindent=1.6em\hangafter=1 Bentivoglio, R., Isufi, E., Jonkman, S. N., \& Taormina, R. (2023). Rapid spatio-temporal flood modelling via hydraulics-based graph neural networks. \emph{Hydrology and Earth System Sciences}, \emph{27}(23), 4227--4246. \url{https://doi.org/10.5194/hess-27-4227-2023}\par\smallskip
\noindent\hangindent=1.6em\hangafter=1 Brandstetter, J., Worrall, D. E., \& Welling, M. (2022). Message passing neural pde solvers. In \emph{International Conference on Learning Representations}. \url{https://openreview.net/forum?id=vSix3HPYKSU}\par\smallskip
\noindent\hangindent=1.6em\hangafter=1 Cheng, Y., Xia, J., Zhou, M., Deng, S., Wang, Z., \& Lu, J. (2023). Parameter sensitivity and uncertainty of a one-dimensional morphodynamic model in the lower yellow river. \emph{Journal of Hydrology}, \emph{625}, 130016. \url{https://doi.org/10.1016/j.jhydrol.2023.130016}\par\smallskip
\noindent\hangindent=1.6em\hangafter=1 Cho, S. W., Lee, J. Y., \& Hwang, H. J. (2026). Learning time-dependent PDE via graph neural networks and deep operator network for robust accuracy on irregular grids. \emph{Journal of Computational Physics}, \emph{544}, 114430. \url{https://doi.org/10.1016/j.jcp.2025.114430}\par\smallskip
\noindent\hangindent=1.6em\hangafter=1 Deal, E., Venditti, J. G., Benavides, S. J., Bradley, R., Zhang, Q., Kamrin, K., \& Perron, J. T. (2023). Grain shape effects in bed load sediment transport. \emph{Nature}, \emph{613}(7943), 298--302. \url{https://doi.org/10.1038/s41586-022-05564-6}\par\smallskip
\noindent\hangindent=1.6em\hangafter=1 Fathi, M. M., Liu, Z., Fernandes, A. M., Hren, M. T., Terry Jr., D. O., Nataraj, C., \& Smith, V. (2026). An integrated deep learning framework enables rapid spatiotemporal morphodynamic predictions toward long-term simulations. \emph{Earth Surface Dynamics}, \emph{14}(2), 313--327. \url{https://doi.org/10.5194/esurf-14-313-2026}\par\smallskip
\noindent\hangindent=1.6em\hangafter=1 Foda, A., Abdelhalim, A., \& Elkhedr, M. (2025). Assessing sensitivity of stream migration at foothill areas: Hydrological modeling and spatial analysis of the red sea coastal stream. \emph{Ecological Indicators}, \emph{173}, 113340. \url{https://doi.org/10.1016/j.ecolind.2025.113340}\par\smallskip
\noindent\hangindent=1.6em\hangafter=1 Fraehr, N., Wang, Q. J., Wu, W., \& Nathan, R. (2022). Upskilling low-fidelity hydrodynamic models of flood inundation through spatial analysis and gaussian process learning. \emph{Water Resources Research}, \emph{58}(8), e2022WR032248. \url{https://doi.org/10.1029/2022WR032248}\par\smallskip
\noindent\hangindent=1.6em\hangafter=1 Goldstein, E. B., Coco, G., \& Plant, N. G. (2019). A review of machine learning applications to coastal sediment transport and morphodynamics. \emph{Earth-Science Reviews}, \emph{194}, 97--108. \url{https://doi.org/10.1016/j.earscirev.2019.04.022}\par\smallskip
\noindent\hangindent=1.6em\hangafter=1 Grass, A. J. (1981). \emph{Sediment transport by waves and currents} (Report No. FL29). SERC London Centre for Marine Technology, University College London.\par\smallskip
\noindent\hangindent=1.6em\hangafter=1 Haces-Garcia, F., Ross, N., Glennie, C. L., Rifai, H. S., Hoskere, V., \& Ekhtari, N. (2025). Rapid 2D hydrodynamic flood modeling using deep learning surrogates. \emph{Journal of Hydrology}, \emph{651}, 132561. \url{https://doi.org/10.1016/j.jhydrol.2024.132561}\par\smallskip
\noindent\hangindent=1.6em\hangafter=1 Hosseiny, H., Masteller, C. C., Dale, J. E., \& Phillips, C. B. (2023). Development of a machine learning model for river bed load. \emph{Earth Surface Dynamics}, \emph{11}(4), 681--693. \url{https://doi.org/10.5194/esurf-11-681-2023}\par\smallskip
\noindent\hangindent=1.6em\hangafter=1 Hu, P., Lyu, B., Li, J., Sun, M., Li, W., \& Cao, Z. (2022). A new two-phase shallow water hydro-sediment-morphodynamic model based on the HLLC solver and the hybrid LTS/GMaTS approach. \emph{Advances in Water Resources}, \emph{166}, 104254. \url{https://doi.org/10.1016/j.advwatres.2022.104254}\par\smallskip
\noindent\hangindent=1.6em\hangafter=1 Hudson, J., \& Sweby, P. K. (2005). A high-resolution scheme for the equations governing 2D bed-load sediment transport. \emph{International Journal for Numerical Methods in Fluids}, \emph{47}(10--11), 1085--1091. \url{https://doi.org/10.1002/fld.853}\par\smallskip
\noindent\hangindent=1.6em\hangafter=1 Khodakarami, S., Oommen, V., Bora, A., \& Karniadakis, G. E. (2025, March 17). Mitigating spectral bias in neural operators via high-frequency scaling for physical systems. arXiv. \url{https://doi.org/10.48550/arXiv.2503.13695}\par\smallskip
\noindent\hangindent=1.6em\hangafter=1 Kochkov, D., Smith, J. A., Alieva, A., Wang, Q., Brenner, M. P., \& Hoyer, S. (2021). Machine learning accelerated computational fluid dynamics. \emph{Proceedings of the National Academy of Sciences}, \emph{118}(21), e2101784118. \url{https://doi.org/10.1073/pnas.2101784118}\par\smallskip
\noindent\hangindent=1.6em\hangafter=1 Kontolati, K., Goswami, S., Em Karniadakis, G., \& Shields, M. D. (2024). Learning nonlinear operators in latent spaces for real-time predictions of complex dynamics in physical systems. \emph{Nature Communications}, \emph{15}(1), 5101. \url{https://doi.org/10.1038/s41467-024-49411-w}\par\smallskip
\noindent\hangindent=1.6em\hangafter=1 Kossaifi, J., Kovachki, N., Li, Z., Pitt, D., Liu-Schiaffini, M., George, R. J., Bonev, B., Azizzadenesheli, K., Berner, J., Duruisseaux, V., \& Anandkumar, A. (2026). A library for learning neural operators. \emph{Journal of Machine Learning Research}, \emph{27}(205), 1--6. \url{https://jmlr.org/papers/v27/26-0434.html}\par\smallskip
\noindent\hangindent=1.6em\hangafter=1 Kovachki, N., Li, Z., Liu, B., Azizzadenesheli, K., Bhattacharya, K., \& Stuart, A. (2023). Neural operator: learning maps between function spaces with applications to PDEs. \emph{Journal of Machine Learning Research}, \emph{24}(1), 4061--4157. \url{https://doi.org/https://doi.org/10.48550/arXiv.2108.08481}\par\smallskip
\noindent\hangindent=1.6em\hangafter=1 Kroon, A., Christiaanse, J. C., Luijendijk, A. P., Schipper, M. A. D., \& Ranasinghe, R. (2025). Parameter uncertainty in medium-term coastal morphodynamic modeling. \emph{Scientific Reports}, \emph{15}(1), 18471. \url{https://doi.org/10.1038/s41598-025-02300-8}\par\smallskip
\noindent\hangindent=1.6em\hangafter=1 Li, Z., Kovachki, N., Azizzadenesheli, K., Liu, B., Bhattacharya, K., Stuart, A., \& Anandkumar, A. (2020). Neural operator: graph kernel network for partial differential equations. \emph{arXiv Preprint arXiv:2003.03485}, 1--21. \url{https://doi.org/10.48550/arXiv.2003.03485}\par\smallskip
\noindent\hangindent=1.6em\hangafter=1 Li, Z., Kovachki, N., Azizzadenesheli, K., Liu, B., Bhattacharya, K., Stuart, A., \& Anandkumar, A. (2021, May 17). Fourier neural operator for parametric partial differential equations. arXiv. \url{https://doi.org/10.48550/arXiv.2010.08895}\par\smallskip
\noindent\hangindent=1.6em\hangafter=1 Li, Z., Kovachki, N. B., Choy, C., Li, B., Kossaifi, J., Otta, S. P., et al. (2023). Geometry-informed neural operator for large-scale 3D PDEs (Vol. 36, pp. 35836--35854). Presented at the Advances in Neural Information Processing Systems, Curran Associates, Inc. \url{https://doi.org/10.52202/075280-1556}.\par\smallskip
\noindent\hangindent=1.6em\hangafter=1 Liu, W., Wu, Q., Liu, Z., Zheng, X., Liu, T., \& Gao, H. (2026). PINN framework for urban flood depth prediction: Integrating data-driven insights with physical constraints. \emph{Journal of Hydrology}, \emph{665}, 134693. \url{https://doi.org/10.1016/j.jhydrol.2025.134693}\par\smallskip
\noindent\hangindent=1.6em\hangafter=1 Liu, X., Song, Y., \& Shen, C. (2024). Bathymetry inversion using a deep-learning-based surrogate for shallow water equations solvers. \emph{Water Resources Research}, \emph{60}(3), e2023WR035890. \url{https://doi.org/10.1029/2023WR035890}\par\smallskip
\noindent\hangindent=1.6em\hangafter=1 Lu, L., Jin, P., Pang, G., Zhang, Z., \& Karniadakis, G. E. (2021). Learning nonlinear operators via DeepONet based on the universal approximation theorem of operators. \emph{Nature Machine Intelligence}, \emph{3}(3), 218--229. \url{https://doi.org/10.1038/s42256-021-00302-5}\par\smallskip
\noindent\hangindent=1.6em\hangafter=1 Mayerle, R., Narayanan, R., Etri, T., \& Abd Wahab, A. K. (2015). A case study of sediment transport in the paranagua estuary complex in brazil. \emph{Ocean Engineering}, \emph{106}, 161--174. \url{https://doi.org/10.1016/j.oceaneng.2015.06.025}\par\smallskip
\noindent\hangindent=1.6em\hangafter=1 Mir, A. A., Patel, M., Albalawi, F., Bajaj, M., \& Tuka, M. B. (2024). A comparative ensemble approach to bedload prediction using metaheuristic machine learning. \emph{Scientific Reports}, \emph{14}(1), 25725. \url{https://doi.org/10.1038/s41598-024-75118-5}\par\smallskip
\noindent\hangindent=1.6em\hangafter=1 Morgan, J. A., Kumar, N., Horner-Devine, A. R., Ahrendt, S., Istanbullouglu, E., \& Bandaragoda, C. (2020). The use of a morphological acceleration factor in the simulation of large-scale fluvial morphodynamics. \emph{Geomorphology}, \emph{356}, 107088. \url{https://doi.org/10.1016/j.geomorph.2020.107088}\par\smallskip
\noindent\hangindent=1.6em\hangafter=1 Mousavi, S., Wen, S., Lingsch, L., Herde, M., Raoni\'c, B., \& Mishra, S. (2025, October 14). RIGNO: a graph-based framework for robust and accurate operator learning for PDEs on arbitrary domains. arXiv. \url{https://doi.org/10.48550/arXiv.2501.19205}\par\smallskip
\noindent\hangindent=1.6em\hangafter=1 Paola, C., \& Voller, V. R. (2005). A generalized exner equation for sediment mass balance. \emph{Journal of Geophysical Research}, \emph{110}(F4), 1--8. \url{https://doi.org/10.1029/2004JF000274}\par\smallskip
\noindent\hangindent=1.6em\hangafter=1 Paul, A., \& Biswas, M. (2019). Changes in river bed terrain and its impact on flood propagation -- a case study of river jayanti, west bengal, india. \emph{Geomatics, Natural Hazards And Risk}, \emph{10}(1), 1928--1947. \url{https://doi.org/10.1080/19475705.2019.1650124}\par\smallskip
\noindent\hangindent=1.6em\hangafter=1 Pfaff, T., Fortunato, M., Sanchez-Gonzalez, A., \& Battaglia, P. W. (2021, June 18). Learning mesh-based simulation with graph networks. arXiv. \url{https://doi.org/10.48550/arXiv.2010.03409}\par\smallskip
\noindent\hangindent=1.6em\hangafter=1 Phillip Lippe, Veeling, B. S., Perdikaris, P., Turner, R. E., \& Brandstetter, J. (2023). PDE-refiner: Achieving accurate long rollouts with neural PDE solvers. In \emph{Advances in Neural Information Processing Systems} (Vol. 36, pp. 1--31). New Orleans, LA, USA: Curran Associates, Inc. \url{https://doi.org/10.52202/075280-2946}.\par\smallskip
\noindent\hangindent=1.6em\hangafter=1 Pourzangbar, A., Oberle, P., Kron, A., \& J. Franca, M. (2024). Analysis of the utilization of machine learning to map flood susceptibility. SSRN. \url{https://doi.org/10.2139/ssrn.4857028}\par\smallskip
\noindent\hangindent=1.6em\hangafter=1 Qi, X., De Almeida, G. A. M., \& Maldonado, S. (2024). Physics-informed neural networks for solving flow problems modeled by the 2D shallow water equations without labeled data. \emph{Journal of Hydrology}, \emph{636}, 131263. \url{https://doi.org/10.1016/j.jhydrol.2024.131263}\par\smallskip
\noindent\hangindent=1.6em\hangafter=1 Rahaman, N., Baratin, A., Arpit, D., Draxler, F., Lin, M., Hamprecht, F. A., et al. (2019, May 31). On the spectral bias of neural networks. arXiv. \url{https://doi.org/10.48550/arXiv.1806.08734}\par\smallskip
\noindent\hangindent=1.6em\hangafter=1 Reyns, J., Jagers, B., Ranasinghe, R., Kernkamp, H., \& Roelvink, D. (2026). A three-dimensional process-based morphodynamic model on orthogonal unstructured grids for coastal applications. \emph{Coastal Engineering}, \emph{204}, 104915. \url{https://doi.org/10.1016/j.coastaleng.2025.104915}\par\smallskip
\noindent\hangindent=1.6em\hangafter=1 Sun, A. Y., Jiang, P., Mudunuru, M. K., \& Chen, X. (2021). Explore spatio-temporal learning of large sample hydrology using graph neural networks. \emph{Water Resources Research}, \emph{57}(12), e2021WR030394. \url{https://doi.org/10.1029/2021WR030394}\par\smallskip
\noindent\hangindent=1.6em\hangafter=1 Taghizadeh, M., Zandsalimi, Z., Nabian, M. A., Shafiee-Jood, M., \& Alemazkoor, N. (2025). Interpretable physics-informed graph neural networks for flood forecasting. \emph{Computer-Aided Civil and Infrastructure Engineering}, \emph{40}(18), 2629--2649. \url{https://doi.org/10.1111/mice.13484}\par\smallskip
\noindent\hangindent=1.6em\hangafter=1 Tassi, P., Benson, T., Delinares, M., Fontaine, J., Huybrechts, N., Kopmann, R., et al. (2023). GAIA - a unified framework for sediment transport and bed evolution in rivers, coastal seas and transitional waters in the TELEMAC-MASCARET modelling system. \emph{Environmental Modelling and Software}, \emph{159}, 105544. \url{https://doi.org/10.1016/j.envsoft.2022.105544}\par\smallskip
\noindent\hangindent=1.6em\hangafter=1 Tipu, R. K., Bhakhar, R., Pandya, K. S., \& Panchal, V. R. (2025). Physics-informed neural networks for predicting sediment transport in pressurized pipe flows. \emph{Environmental Earth Sciences}, \emph{84}(11), 292--318. \url{https://doi.org/10.1007/s12665-025-12295-0}\par\smallskip
\noindent\hangindent=1.6em\hangafter=1 Vanzo, D., Peter, S., Vonwiller, L., B\"urgler, M., Weberndorfer, M., Siviglia, A., et al. (2021). basement v3: A modular freeware for river process modelling over multiple computational backends. \emph{Environmental Modelling and Software}, \emph{143}, 105102. \url{https://doi.org/10.1016/j.envsoft.2021.105102}\par\smallskip
\noindent\hangindent=1.6em\hangafter=1 Vriend, H. J. D. (1987). 2DH mathematical modelling of morphological evolutions in shallow water. \emph{Coastal Engineering}, \emph{11}(1), 1--27. \url{https://doi.org/10.1016/0378-3839(87)90037-8}\par\smallskip
\noindent\hangindent=1.6em\hangafter=1 Vundavilli, H., Mullarney, J. C., MacDonald, I. T., \& Bryan, K. R. (2021). The interaction of buoyant coastal river plumes with mangrove vegetation and consequences for sediment deposition and erosion in a tidal environment. \emph{Continental Shelf Research}, \emph{222}, 104417. \url{https://doi.org/10.1016/j.csr.2021.104417}\par\smallskip
\noindent\hangindent=1.6em\hangafter=1 Vundavilli, H., Mullarney, J. C., \& MacDonald, I. T. (2024). The coalescence of two mangrove-lined river plumes and consequences for sediment transport and deposition in a coastal environment. \emph{Continental Shelf Research}, \emph{279}, 105280. \url{https://doi.org/10.1016/j.csr.2024.105280}\par\smallskip
\noindent\hangindent=1.6em\hangafter=1 Wu, Z., Yang, Y., Xia, J., Xu, Z., Shao, D., \& Xu, W. (2026). Re-thinking levee strategies for aggradation-driven flood risk in the waiho river, new zealand. \emph{Journal of Hydrology: Regional Studies}, \emph{64}, 103143. \url{https://doi.org/10.1016/j.ejrh.2026.103143}\par\smallskip
\noindent\hangindent=1.6em\hangafter=1 Yen, C., \& Lee, K. T. (1995). Bed topography and sediment sorting in channel bend with unsteady flow. \emph{Journal of Hydraulic Engineering}, \emph{121}(8), 591--599. \url{https://doi.org/10.1061/(ASCE)0733-9429(1995)121:8(591)}\par\smallskip
\noindent\hangindent=1.6em\hangafter=1 Yu, Q., Rennie, C. D., \& Slaney, J. M. (2025). Gravel-bed river morphodynamic processes throughout a large flood with an 80-year return period from numerical modeling: implications for flood regulation strategies. \emph{Journal of Hydrology}, \emph{660}, 133463. \url{https://doi.org/10.1016/j.jhydrol.2025.133463}\par\smallskip
\noindent\hangindent=1.6em\hangafter=1 Zhao, S., Li, Z., Fan, B., Wang, Y., Yang, H., \& Wang, J. (2025). LESnets (large-eddy simulation nets): Physics-informed neural operator for large-eddy simulation of turbulence. \emph{Journal of Computational Physics}, \emph{537}, 114125. \url{https://doi.org/10.1016/j.jcp.2025.114125}\par\smallskip
\endgroup
\end{document}